\documentclass[12pt,reqno]{amsart}

\usepackage[centertags]{amsmath}
\usepackage{amsfonts}
\usepackage{amssymb}
\usepackage{amsthm}
\usepackage{newlfont}

\newtheorem{thm}{Theorem}[section]
\newtheorem{cor}[thm]{Corollary}
\newtheorem{lem}[thm]{Lemma}
\newtheorem{prop}[thm]{Proposition}

\theoremstyle{definition}

\theoremstyle{remark}
\newtheorem{rem}[thm]{Remark}
\numberwithin{equation}{section}

\newcommand{\C}{\mathbb{C}}
\newcommand{\R}{\mathbb{R}}
\newcommand{\N}{\mathbb{N}}
\newcommand{\Z}{\mathbb{Z}}

\newcommand{\mfr}[1]{\mathfrak{#1}}
\newcommand{\frC}{\mfr{C}}

\newcommand{\dif}{\partial}
\newcommand{\md}{d\kern-0.035cm\char39\kern-0.03cm}
\newcommand{\bsl}{\backslash}
\newcommand{\medz}{\vspace{2mm}}
\newcommand{\ol}{\overline}

\newcommand{\itt}{\intertext}
\newcommand{\dL}{\dif\La}
\newcommand{\Pro}{\mathsf{P}}
\newcommand{\Exp}{\mathsf{E}}
\newcommand{\Dis}{\mathsf{F}}
\newcommand{\non}{\nonumber}

\newcommand{\ii}{\'\i}
\newcommand{\ie}{i.e.\ }
\newcommand{\eg}{e.g.\ }

\newcommand{\cf}{c.f.\ }

 \DeclareMathOperator{\Inn}{Int}
 \DeclareMathOperator{\Ext}{Ext}

\DeclareMathOperator{\co}{con} 
\DeclareMathOperator{\card}{card}

\let\al=\alpha \let\be=\beta \let\de=\delta \let\ep=\epsilon
\let\ve=\varepsilon \let\vp=\varphi \let\ga=\gamma \let\io=\iota
\let\ka=\kappa \let\la=\lambda \let\om=\omega \let\vr=\varrho
\let\si=\sigma  \let\th=\theta \let\vt=\vartheta
\let\ze=\zeta   \let\vk=\varkappa

\let\De=\Delta \let\Ga=\Gamma \let\La=\Lambda \let\Om=\Omega
\let\Th=\Theta  

\newcommand{\frL}{\mathfrak{L}}
\newcommand{\frY}{\mathfrak{Y}}

\newcommand{\cA}{\mathcal{A}} 
\newcommand{\cC}{\mathcal{C}} \newcommand{\cD}{\mathcal{D}}
\newcommand{\cE}{\mathcal{E}} 
\newcommand{\cG}{\mathcal{G}} 
\newcommand{\cI}{\mathcal{I}} 
\newcommand{\cK}{\mathcal{K}} 
 
\newcommand{\cO}{\mathcal{O}} \newcommand{\cP}{\mathcal{P}}

 \newcommand{\cY}{\mathcal{Y}}
\newcommand{\cZ}{\mathcal{Z}}

\begin{document}

\title[Chaotic size dependence in the Ising]{Chaotic size
       dependence in the Ising model with random boundary
       conditions}%

\author{A.~C.~D.~van Enter}%
\address{Institute for Theoretical Physics \\ University of
  Groningen \\ Groningen \\ The Netherlands}%
\email{A.C.D.van.Enter@phys.rug.nl}%

\author{I.~Medve\md}%
\address{Department of Chemistry \\ Texas Christian University \\
  Fort Worth \\ USA}%
\email{I.Medved@tcu.edu}%

\author{K.~Neto\v cn\'y}%
\address{Institute for Theoretical Physics \\ Catholic University
  of Leuven \\ Leuven \\ Belgium}%
\email{Karel.Netocny@fys.kuleuven.ac.be}%

\subjclass{82B20, 82B44, 60F05, 60K35}

\keywords{Random boundary conditions, metastates, contour models,
local- and central-limit theorems.}

\begin{abstract}
We study the nearest-neighbour Ising model with a class of random
boundary conditions, chosen from a symmetric i.i.d.\ distribution.
We show for dimensions 4 and higher that almost surely the only
limit points for a sequence of increasing cubes are the plus and
the minus state. For d=2 and d=3 we prove a similar result for
sparse sequences of increasing cubes. This question was raised by
Newman and Stein. Our results imply that the Newman-Stein
metastate is concentrated on the plus and the minus state.
\end{abstract}

\maketitle

\section{Introduction}

In most studies of phase transitions, one considers boundary
conditions which are either symmetric between the different
possible phases (such as free, periodic or antiperiodic boundary
conditions in low temperature Ising, Potts or vector models) or a
priori known to be in some sense typical for one of the phases.
The latter case in its best-known (and purest) form is realized by
the choice of plus or minus boundary conditions for the Ising
model, and in some sense by the wired or free boundary conditions
for the random cluster model.  Other examples of an unambiguous
preference for one phase are the cases of uniform weak boundary
conditions \cite{LebPen, Med01} for low temperature Ising or Potts
models, predominantly plus boundary conditions for the Ising model
\cite{Higyos}, and random or weak boundary conditions at the
high-$q$ Potts transition temperature \cite{Campent,Med01,vE00}.
Complementary results generalizing (deterministic) symmetric
boundary conditions can be found in \cite{BK90, BKM91, BK95,
Med01}, for instance.

A question which comes up naturally in the theory of spin-glasses,
where the phases are unknown so that the choice of coherent
boundary conditions is not available (although one can put the
question in substantial greater generality), is what will happen
if one chooses the boundary conditions randomly, without having a
systematic preference for one of the phases.

This is the type of question we want to study here. To be more
definite, we consider the standard Ising model at low temperature,
with symmetric i.i.d.~boundary conditions, \cf~\cite{NeSt92},
example iii.4. It is conjectured there (compare also \cite{FH,
vE90}) that when one would take an increasing sequence of volumes,
one would oscillate randomly between being close to the plus and
the minus phase. Heuristically, the fluctuations of the free
energies of the plus and minus states should scale with the square
root of the boundary, which diverges with increasing volumes.
Thus, one would expect that weight of either the plus or the minus
measure in a mixture will be of order $e^{- V^{(d-1)/2d}}$, which
becomes negligible for sufficiently large volumes.

This non-convergence to a single thermodynamic limit measure is an
example of what Newman and Stein call ``chaotic size dependence''.
They have developed their ideas within the formalism of metastates
\cite{NeSt97,NeSt98,Ne97,NeSt02}. Since then similar arguments
have been made precise for a class of random mean-field models
\cite{BovGay,Kue97,Kue98,Kue98pr,Boven,vES02}.  However, for
short-range models hardly any precise result has been obtained.

Here we study a simple version of this problem for the standard
Ising model. Our simplification is the condition that the bonds on
the boundary are weaker than the bonds in the bulk. This removes
``by hand'' any interface (large contour) because these get
attracted to the boundary. Our result is that the above heuristics
is correct, and that a chaotic size dependence occurs indeed. The
two possible limit states occurring with the same limit frequency
are the plus and the minus states. In other words, the metastate
is concentrated with equal weight on these two pure states. This
can be shown for a sequence of strictly increasing cubes in high
dimensions, in dimension 2 and 3 we need to impose an extra
condition of ``sparsity'' on the  sequence of increasing volumes
(c.f.~Corollary~\ref{cor: metastates}).

The structure of the set of limit states can be inferred through a
toy-model, where only two configurations are allowed (the
zero-temperature approximation). Having in mind the Ising model
coupled to a random environment, let $\La_n\subset\Z^d$ be a cube
with side $n$ and $\si \in \{-1,1\}$ the possible states.
Considering a collection $\{\la_x\}_{x\in\Z^d}$ of identical
independent variables with the distribution $\Pro(-1) = \Pro(1) =
\frac12$, the Hamiltonian of the toy-model is
\begin{equation}
  H^{\la}_n(\si) = - \si \sum_{x \in \dif\La_n} \la_x\ .
\end{equation}
Using $\mu_{n}^{\la}$ to denote the corresponding finite-volume
Gibbs measure, we immediately have $\mu_{n}^{\la}(\si) = \tanh
S_{n}^{\la}$ with $S_{n}^{\la} =  \sum_{x \in \dL_n} \la_x$. Thus,
in order to find the limit points of $\{\mu_{n}^{\la}\}$, it is
sufficient to find the limit points of $\{S_{n}^{\la}\}$. However,
the latter is a sequence of ``essentially'' independent functions,
each of them being the sum of independent variables. Therefore,
one can readily use the local-limit theorem to show that
$\Pro(S_{n}^{\la} = k) \simeq n^{-\frac{d-1}2}$ for any $k \in
2\Z$. Then, according to the Borel-Cantelli lemmas, $k \in 2\Z$ is
a limit point of $\{S_{n}^{\la}\}$ $\Pro$-a.s.\ if and only if
$\sum_{n} \Pro(S_{n}^{\la} = k) = \infty$. Realizing that
$-\infty$ and $\infty$ are always limit points $\Pro$-a.s.\ (due
to the Borel-Cantelli argument and the symmetry of the
distribution), we can conclude the following: for $d = 2,3$ the
set of limit points of $\{S_{n}^{\la}\}$ is $2\Z \cup
\{-\infty,\infty\}$ $\Pro$-a.s., while for $d > 3$ it is
$\{-\infty,\infty\}$ $\Pro$-a.s. As a consequence, the only limit
points of the sequence of the Gibbs measures for $d > 3$ are
$\de_{1}$ and $\de_{-1}$ $\Pro$-a.s. On the other hand, in
dimensions $d = 2,3$ we obtain an infinite set of limit measures.
The set of mixed limit measures, however, is null-recurrent. Note
that this picture differs from the case of free boundary
conditions, where there is only one limit state
$\frac{1}{2}(\de_{1} + \de_{-1})$.

In the sequel, we show that this behaviour is stable with respect
to thermal fluctuations for $d > 3$. This is done by generalizing
the above scenario and proving a weak variant of the local-limit
theorem for the boundary term. A similar result is obtained in
dimensions 2 and 3 for ``sparse'' sequences. Although this
behaviour seems almost ``physically obvious'' (as one can see from
the toy model), the proof turns out, somewhat surprisingly, to be
rather non-trivial.

On the other hand, the full low-dimensional structure of limit
points remains out of the scope of the present paper. Presumably,
mixed states can appear as limit points, but again null recurrent,
that is, with disappearing probabilities.

The organization of the paper is as follows. In Section~\ref{sec:
model&results}, we specify the studied model and state our main
result in Theorem~\ref{thm: main}. Section~\ref{sec: contour rep.}
is devoted to the contour analysis of our model, yielding its
reformulation in terms of contour ensembles. The cluster-expansion
control over these ensembles is provided in Section~\ref{sec:
perturbative control}. This leads to Proposition~\ref{prop:
equivalence of measures} and its Corollary~\ref{cor: equivalence}
from which we can conclude that no long contours will appear
inside the system. The proof of Theorem~\ref{thm: main} is then
finished by the probabilistic arguments of Section~\ref{sec:
prob.anal.}. We again employ cluster-expansion techniques, now to
prove a weak version of the local limit theorem (Lemma~\ref{lem:
weak LLT}). A Borel-Cantelli argument (Proposition 5.4) then
closes the argument. Some comments and remarks are given in
Section~\ref{sec: concl. remarks}. Technicalities concerning
statements for the abstract polymer model and their application to
the proof of the convergence of cluster expansions needed in
Sections~\ref{sec: perturbative control}~and~\ref{sec: prob.anal.}
are deferred to the appendices.

\section{Model and Results}\label{sec: model&results}

Let $\{\La_n\}_{n\in\N}$ be the sequence of $d$-dimensional cubes
on $\Z^d$, $d\geq 2$, given by\footnote{Notice that the cube
$\La_n$ has side-length $n-1$ and is centred at the origin if $n$
is odd and at $\frac12\,\mathbb{I}$ if $n$ is even, where
$\mathbb{I}\in\Z^d$ is the unit vector.}
\begin{equation}
   \La_n = \{ x\in\Z^d \,:\,  -\;\frac n2 < x_i \leq \frac n2
                 \quad \forall \; i=1,\dots,d \} .
\end{equation}
In this paper, we study the ferromagnetic, nearest-neighbour Ising
model in $\La_n$ exposed to random boundary fields $\la \in
\R^{\Z^d}$ in the thermodynamic limit $n\to\infty$. Namely, using
$\Om$ to denote the set $\{-1,1\}^{\Z^d}$ of spin configurations
on $\Z^d$ and $\Om_n$ to denote the set $\{-1,1\}^{\La_n}$ of spin
configurations in $\La_n$, we consider the \emph{Hamiltonian}
\begin{equation}\label{eq: Hamiltonian}
   H_n^\la (\si_n) =
   - \; \be \sum_{\langle x,y \rangle \atop x,y\in\La_n}
           (\si_x \si_y -1) \;
   - \; \sum_{x \in \dL_n} \la_x \si_x , \qquad \si_n\in\Om_n.
\end{equation}
Here $\langle x,y\rangle$ stands for a pair of nearest-neighbour
sites $x,y\in\Z^d$, the bulk coupling $\be>0$, the set $\dL_n$
contains all $x\in\La_n$ having at least one nearest-neighbour
site in $\La_n^c$, and $\{\la_x\}_{x\in\Z^d}$ are identical,
independent, symmetrically distributed  random variables with zero
mean. The latter represent random boundary conditions with
boundary terms of strength $\la_x$. We will write $\Pro$ for the
(product) \emph{probability law} of $\la$ and $\Exp$ for the
\emph{expectation} with respect to $\Pro$. Let $\vp(t)=\Exp\,
e^{it\la_0}$, $t\in\R$, be the \emph{characteristic function} of
$\la_0$. We restrict ourselves to distributions with bounded
supports, precisely, we assume that $\Pro(|\la_0|>\la^*) =0$ for a
certain finite $\la^*$.

The finite-volume \emph{Gibbs measure} $\mu_n^\la$ corresponding to the
Hamiltonian
\eqref{eq: Hamiltonian} is defined as
\begin{equation}
    \mu_n^\la (A) = \sum_{\si_n\in A}
       \frac{e^{- H_n^\la (\si_n)}}{Z_n^\la} \,, \quad A\subset\Om_n,
\end{equation}
where the normalizing constant $Z_n^\la = \sum_{\si_n\in\Om_n}
e^{-H_n^\la (\si_n)}$ is the \emph{partition function}. Given
$\be$ sufficiently large, our aim is the analysis of the set of
limit points of the sequence of random measures
$\{\mu_n^\la\}_{n\in\N}$. Let
\begin{gather}
   \mu_n^\pm (A) =  \sum_{\si_n\in A}
       \frac{e^{- H_n^\pm (\si_n)}}{Z_n^\pm} \,, \quad A\subset\Om_n,
        \\ \intertext{where}
   H_n^\pm (\si_n) =
   - \; \be \sum_{\langle x,y \rangle \atop x,y\in\La_n}
           (\si_x \si_y -1) \;
   \mp \; \be \sum_{x \in \dL_n} \si_x
\end{gather}
and $Z_n^\pm = \sum_{\si_n\in\Om_n} e^{-H_n^\pm(\si_{n})}$. The
weak limits $\mu^\pm$ of the sequences $\{\mu_n^\pm\}$ are the
only extremal translation-invariant Gibbs measures of the Ising
model and $\mu^+ \neq \mu^-$ for $d \geq 2$ and $\beta$ large
enough, see \eg \cite{Ge}. In this paper we prove the following
theorem.

\begin{thm}\label{thm: main}
Given $d \geq 2$ and $0 < \la^* < \infty$, there exists a constant
$\be_0=\be_0(\la^*,d) < \infty$ such that for any $\be\geq\be_0$
and any symmetric distribution $\Pro$ of boundary fields with zero
mean, strictly positive variance, and satisfying
$\Pro(|\la_0|>\la^*)=0$, one has:
\begin{enumerate}
\item If $d>3$, then the set of limit points of $\{\mu_n^\la\}_{n\in\N}$ is
      $\{\mu^+, \mu^-\}$ $\Pro$-a.s.
\item If $d\in\{2,3\}$ and $\om>0$, then the set of limit points of the
      ``sparse'' sequence $\{\mu_{[n^{4-d+\om}]}^\la\}_{n\in\N}$ is
      $\{\mu^+, \mu^-\}$ $\Pro$-a.s.
\end{enumerate}
\end{thm}

The conclusion of the theorem implies that the Newman-Stein metastate is
$ \frac{1}{2} (\delta_{\mu^+} +\delta_{\mu^-})$.

The proof of the theorem is carried out in two steps. First, in
Section~\ref{sec: contour rep.} we rewrite our model in terms of
contours, using two auxiliary contour ensembles with the
corresponding measures $\mu_n^{\la,+}$ and $\mu_n^{\la,-}$ and the
partition functions $Z_n^{\la,+}$ and $Z_n^{\la,-}$. This enables
us to express $\mu_n^\la$ through $\mu_n^{\la,+}$,
$\mu_n^{\la,-}$, and $F_n^\la = \log Z_n^{\la,+} - \log
Z_n^{\la,-}$. In Section~\ref{sec: perturbative control} we in
particular show that the occurrence of long contours is excluded
in the region $\be \geq \be_0(\la^*,d)$, where the low-temperature
cluster expansions for $Z_n^{\la,+}$ and $Z_n^{\la,-}$ converge.
Establishing the relation between the limit points of
$\{\mu_n^{\la}\}$ and the limits of $\{\mu_n^{\pm}\}$ in
Corollary~\ref{cor: equivalence}, the original problem gets
reduced to the task of finding the limit points of $\{F_n^\la\}$.
Second, in Section~\ref{sec: prob.anal.} we solve this task by
using probabilistic arguments that have again the structure of a
cluster expansion, this time for a kind of imaginary boundary free
energy. We prove a weak variant of the local-limit theorem which
suffices for our purposes; this is to exclude the occurrence of
a``mixed state'' as a possible limit measure. The structure of the
set of limit points of $\{F_n^\la\}$ is finally obtained with the
help of the Borel-Cantelli lemmas, and is stated in
Proposition~\ref{prop: limit points of F}.

\section{Contour Representations}\label{sec: contour rep.}

In this section, we introduce contour representations for our
model given by the Hamiltonian $H_n^\la$ as well as for the models
corresponding to the Hamiltonians $H_n^+$ and $H_n^-$. In the
former case, we are interested in boundary fields $\la$ of small
strength. Hence, we make use of contours suitable for the study of
lattice models under free boundary conditions. These may be
``open'' and their definition is in the spirit of \cite{BK95}. In
the latter case, the standard, ``closed'' Ising contours are
employed. It turns out that the difference between the two cases
merely concerns ``boundary contours'', \ie those containing the
sites from $\dL_n$, see below.

Our contour representations are set up to allow us to establish
estimates that are uniform in a large class of boundary
conditions. Whenever these estimates hold, long contours will not
appear in typical configurations, and the expansions given in the
next section will converge. When this happens, we can consequently
conclude that we either have a ``typical plus'' or a ``typical
minus'' configuration. We will estimate these two sets of
configurations separately, uniformly in our chosen class of
boundary conditions.

We shall proceed in a slightly more general context, allowing at
the same time to study the expectation of local observables in the
above models as $n\to\infty$. By virtue of an ``FKG-argument'',
see Corollary~\ref{cor: equivalence}, it is sufficient to control
the infinite-volume expectation of the spin at each site
$x\in\Z^d$. Observing that there exists $n_x < \infty$ such that
$x\in\La_n$ for all $n\geq n_x$, we therefore introduce an
external field $\eta\in\R$ at $x$, \ie for each $x\in\Z^d$ we
consider models with the perturbed Hamiltonians
\begin{equation}
  H_n^{\la,x} (\si_n) = H_n^\la (\si_n) - \eta \, \si_x
     \quad \text{and} \quad
  H_n^{\pm,x} (\si_n) = H_n^\pm (\si_n) - \eta \, \si_x  \,.
\end{equation}
Here $\si_x$ is the restriction of $\si_n$ to $x$ if $n\geq n_x$,
while it is $0$ otherwise. All other quantities associated with
these new models will also have the additional superscript $x$.
The superscript will be suppressed whenever we will be in the
original situation, corresponding to $\eta=0$.

Let $\Box_x$, $x\in\Z^d$, be the closed unit cube in $\R^d$ whose
centre is at $x$ and let $V_n = \cup_{x\in\La_n} \Box_x$. Given
$\si_n\in\Om_n$, we define $V_n^\pm (\si_n) = \cup_{x\in\La_n:\;
\si_x = \pm 1} \Box_x \subset V_n$ as the ``$\pm$ regions''
corresponding to $\si_n$ and $\cD(\si_n)$ as the set of connected
components of $V_n^+ (\si_n) \cap V_n^- (\si_n)$. Thus, the set
$\cD(\si_n)$ represents the connected and mutually disjoint
boundaries separating $V_n^+ (\si_n)$ from $V_n^-(\si_n)$. A
\emph{contour} is any element of the union
$\cD_n=\cup_{\si_n\in\Om_n} \cD(\si_n)$. We write $|\ga|$ for the
number of \emph{plaquettes} (\ie closed $(d-1)$-dimensional faces
of the closed unit cubes) lying in the contour $\ga$.

Next, we will define the \emph{interior} and \emph{exterior} of a
contour $\ga$. For each corner $k=[k_1,\dots,k_d]$ of the box
$V_n$,\footnote{If $n$ is odd, $n=2m+1$, then one obviously has
$|k_i|=m$, $i=1,\dots,d$, for the corner $k$. However, if $n$ is
even, $n=2m$, then $|k_i|$ equals either $m$ or $m-1$.} let us
introduce the ``octant'' associated with $k$ as
\begin{equation}
  \cO_n (k) = \{ x\in\R^d :\; x_i \geq k_i \;\;\text{if}\;\;
    i\in I_-,\;  x_i \leq k_i \;\;\text{if}\;\; i\in I_+ \} ,
\end{equation}
where $i\in I_-$ whenever $y_i \geq k_i$ for all $y\in V_n$, while
$i\in I_+$ whenever $y_i \leq k_i$ for all $y\in V_n$. Notice that
$V_n = \cap_k \, \cO_n (k)$. Two possibilities arise:
\begin{enumerate}
  \item There is a corner $k$ of $V_n$ such that $\ga\cap\dif V_n
        \subset \dif \cO_n(k)$. Then $\Inn\ga$ is the union of
        all finite components of $\cO_n(k)\bsl\ga$ and $\Ext\ga$ is
        $V_n \setminus (\ga \cup \Inn\ga)$.\footnote{Cf.\
        \cite{BK95}. This definition does not depend on the choice
        of $k$ if more corners are possible. This case also covers
        the contours which do not touch the boundary $\dif V_n$.}
  \item There is no corner $k$ of $V_n$ for which $\ga\cap\dif V_n
        \subset \dif \cO_n(k)$, \ie there is a kind of interface. We
        then choose $\Ext\ga$ to be the largest component of
        $V_n\bsl\ga$ and $\Inn\ga$ to be the union of the remaining
        components of $V_n\bsl\ga$.\footnote{
        If there are several components of $V_n\bsl\ga$ with the
        largest volume, we take the first one in some fixed (\eg
        lexicographic) order.}
\end{enumerate}
We point out that the joint exterior $\Ext(\dif)=
\cap_{\ga\in\dif} \Ext\ga$ of any set of contours $\dif\in\cD_n$
such that $\dif=\cD(\si_n)$ is either a subset of $V_n^+ (\si_n)$
or $V_n^- (\si_n)$. Hence, the set $\Om_n$ may be written as a
union $\Om_n^+ \cup \Om_n^-$ of disjoint subsets $\Om_n^\pm = \{
\si_n \in \Om_n:\, \Ext(\cD(\si_n)) \subset V_n^\pm (\si_n)\}$.
Finally, let $\La(\ga)=\Inn\ga\cap\La_n$ and $v(\ga) = \max_{\ga'
\subset \Inn\ga} |\La(\ga')|$ for any contour $\ga\in\cD_n$.

We shall now rewrite the partition function $Z_n^{\la,x}$ in terms
of contours. Given $\Th\subset\Z^d$, let $S_\Th^\la = \sum_{x\in
\Th\cap\dL_n} \la_x$ and $E_\Th^{\la,x,\pm} = \mp( S_\Th^\la +
\eta \mathbf{1}_{x\in\Th})$, $x\in\Z^d$. For simplicity, we write
$S_n^\la = S_{\dL_n}^\la $ and $E_n^{\la,x,\pm} =
E_{\La_n}^{\la,x,\pm}$. Let us introduce the quantities
$Z_\ga^{\la,x,\pm}$ and $K_\ga^{\la,x,\pm}$ for any contour
$\ga\in\cD_n$ in the following inductive manner:
\begin{enumerate}
\item We set
  \begin{equation}\label{eq: Z,K for 1-contour}
     \qquad \qquad
     Z_\ga^{\la,x,\pm} = e^{- E_{\La(\ga)}^{\la,x,\pm} }
     \quad \text{and} \quad
     K_\ga^{\la,x,\pm} = e^{-2\be|\ga|
       + E_{\La(\ga)}^{\la,x,\pm}  -  E_{\La(\ga)}^{\la,x,\mp} }
  \end{equation}
  for any contour $\ga$ with $v(\ga)=0$.
\item Assuming that $K_{\ga'}^{\la,x,\pm}$ and $Z_{\ga'}^{\la,x,\pm}$
  have been defined for all contours having $v(\ga')< N \leq |\La_n|$,
  for any $\ga$ with $v(\ga)=N$ we set
  \begin{gather}\label{eq: Z for N-contour}
     Z_\ga^{\la,x,\pm} = e^{-E_{\La(\ga)}^{\la,x,\pm} }
         \sum_{\dif \sqsubset \Inn\ga} \;
         \prod_{\ga'\in\dif} K_{\ga'}^{\la,x,\pm} \\
     \intertext{and}\label{eq: K for N-contour}
      K_\ga^{\la,x,\pm} = e^{-2\be|\ga|} \;
        \frac{ Z_\ga^{\la,x,\mp}}{Z_\ga^{\la,x,\pm}} \,.
  \end{gather}
  Here the sum is over all families $\dif\subset\cD_n$ of mutually
  disjoint contours which all lie in $\Inn\ga$; the term
  corresponding to $\dif=\emptyset$ is set equal to $1$.
\end{enumerate}
In the second step one uses the fact that $v(\ga')< v(\ga)$ for
any $\ga'$ which lies in $\Inn\ga$ (notice that $\ga\cap\Inn\ga =
\emptyset$). Observing that
\begin{equation}
   H_n^{\la,x} (\si_n) = 2\be \sum_{\ga\in\cD(\si_n)} |\ga|
   + E^{\la,x,+}_{V_n^+ (\si_n)\cap\La_n}
   + E^{\la,x,-}_{V_n^- (\si_n)\cap\La_n}
\end{equation}
for all $\si_n\in\Om_n$, one may use standard arguments \cite{Z84,
BI89, BK95} to express $Z_n^{\la,x}$ as the sum of two partition
functions of auxiliary contour ensembles. Namely, one has
\begin{equation}
   Z_n^{\la,x} = Z_n^{\la,x,+} + Z_n^{\la,x,-}
   \qquad \text{with} \quad
   Z_n^{\la,x,\pm} = e^{-E_n^{\la,x,\pm}}
       \sum_{\dif \sqsubset V_n} \;
       \prod_{\ga\in\dif} K_\ga^{\la,x,\pm} ,
\end{equation}
where the summation is over all families $\dif\subset\cD_n$ of
mutually disjoint contours; the term corresponding to
$\dif=\emptyset$ is set equal to $1$. Each of the contour
ensembles may be associated with a measure $\mu_n^{\la,x,\pm}$
given through the restricted sets of configurations $\Om^\pm_n$,
\begin{equation}
   \mu_n^{\la,x,\pm} = \begin{cases}
      \frac{e^{-H_n^{\la,x} (\si_n)}}{Z_n^{\la,x,\pm}} &
             \si_n\in\Om_n^\pm , \\
      0  & \text{otherwise}.
   \end{cases}
\end{equation}
These contour ensembles provide a suitable representation for the
finite-volume Gibbs measure $\mu_n^{\la}$,
\begin{equation}\label{eq: main relation}
  \mu_n^\la = \frac{Z_n^{\la,+} \mu_n^{\la,+} +
    Z_n^{\la,-} \mu_n^{\la,-}}{Z_n^{\la,+} + Z_n^{\la,-}} =
  \frac{\mu_n^{\la,+}}{1+e^{-F_n^\la}} +
   \frac{\mu_n^{\la,-}}{1+e^{F_n^\la}} \ ,
\end{equation}
where
\begin{equation}
    F_n^\la = \log Z_n^{\la,+}  - \log Z_n^{\la,-}\ .
\end{equation}

In the case of $\pm$ boundary conditions, one considers the set
$\cD^\pm(\si_n)$ of connected components of the boundary $\dif
V_n^\mp (\si_n)$. The set of contours in $V_n$ is then defined as
$\hat\cD_n = \cup_{\si_n\in\Om_n} \cD^+ (\si_n) =
\cup_{\si_n\in\Om_n} \cD^- (\si_n)$. Given $\hat\ga\in\hat\cD_n$,
its interior and exterior are introduced naturally: $\Inn\hat\ga$
is the union of all finite components of $\R^d\setminus\hat\ga$
and $\Ext\hat\ga = V_n \setminus (\hat\ga \cup \Inn\hat\ga)$.
Again, we let $\La(\hat\ga)=\Inn\hat\ga \cap \La_n$ and
$v(\hat\ga) = \max_{\hat\ga' \subset\hat\ga} |\La (\hat\ga')|$.
Clearly, the sets $\cD_n$ and $\hat\cD_n$ only differ in
``boundary contours'' since $\La(\ga) \cap \dL_n = \emptyset$ iff
$\ga \in \cD_n \cap \hat\cD_n$. Setting $E_\Th^{\pm,x} = \mp \eta
\, \mathbf{1}_{x\in\Th}$ for any $\Th\subset\Z^d$ and $x\in\Z^d$
and observing that
\begin{equation}
   H_n^{\pm,x} (\si_n) = 2\be \sum_{\ga\in\cD(\si_n)} |\ga|
   + E^{+,x}_{V_n^+ (\si_n)\cap\La_n}
   + E^{-,x}_{V_n^- (\si_n)\cap\La_n},
\end{equation}
the quantities $Z_{\hat\ga}^{\pm,x}$ and $K_{\hat\ga}^{\pm,x}$ are
introduced in an inductive manner analogously to
$Z_\ga^{\la,x,\pm}$ and $K_\ga^{\la,x,\pm}$. It then follows that
\begin{equation}
   Z_n^{\pm,x} = e^{- E_n^{\pm,x}}
       \sum_{\hat\dif \sqsubset V_n} \;
       \prod_{\hat\ga\in\hat\dif} K_{\hat\ga}^{\pm,x} ,
\end{equation}
where the sum goes over all families $\hat\dif\subset\hat\cD_n$ of
mutually disjoint contours with the term corresponding to
$\hat\dif=\emptyset$ being set equal to $1$.

\section{Perturbative Control of the Contour Ensembles}\label{sec:
perturbative control}

Let us define the relation $\io$ of incompatibility on the sets
$\cD_n$ and $\hat\cD_n$ of contours in $V_n$ as disjointness (for
instance, for any $\ga,\ga' \in \cD_n$ one has $\ga \io \ga'$ iff
$\ga \cap \ga' \neq \emptyset$). Then the logarithms of the
partition functions $Z_n^{\la,x,\pm}$ and $Z_n^{\pm,x}$ as well as
those of $Z_\ga^{\la,x,\pm}$ and $Z_{\hat\ga}^{\pm,x}$ can be
expressed in the form of cluster expansions (c.f.~Appendix
\ref{sec: clust.exp.}). Namely, writing $\cC_n$ and $\hat\cC_n$
for the set of all clusters $C\subset\cD_n$ and $\hat
C\subset\hat\cD_n$, respectively, one has
\begin{gather}\label{eq: expansion Z la}
  \log Z_n^{\la,x,\pm} = - E_n^{\la,x,\pm} +
        \sum_{C\in\cC_n} \Phi_C^{\la,x,\pm}  \\
  \intertext{and} \label{eq: expansion Z +-}
  \log Z_n^{\pm,x} = - E_n^{\pm,x} +
       \sum_{\hat C\in\hat\cC_n} \Phi_{\hat C}^{\pm,x} ,
       \quad x\in\Z^d \,.
\end{gather}
The convergence of these series as well as the convergence their
derivatives with respect to $\eta$ is guaranteed by the following
two lemmas. Their proofs, carried out with the help of the
Koteck\'y-Preiss criterion \eqref{eq: KP-general} (see \cite{KoPr}
and also \cite{BoZa,MS00,Sok01}, for instance), are deferred to
Appendix~\ref{sec: proof of cl.exp.}.


\begin{lem}\label{Lem: cl.exp.}
Let $d\geq 2$ and $\la^*,\eta^*\geq 0$. There exist
$c_1,c_2<\tau<\infty$ such that for any $\be\geq\tau$ and
$x_0\in\Z^d$ one has
\begin{gather}\label{eq: phi-decay}
   \sum_{C:\, \La(C)\ni x} e^{2(\be-c_1)|C|}
      |\Phi^{\la,x_0,\pm}_C| \leq 1 \\
   \itt{and} \label{eq: phi'-decay}
   \sum_{C:\, \La(C)\ni x} e^{2(\be-c_2) |C|} \Bigl|
         \frac{\dif\Phi^{\la,x_0,\pm}_C}{\dif\eta} \Bigr|
       \leq 1
\end{gather}
for all $x\in\La_n$ and $n\in\N$ whenever $|\la_y|\leq\la^*$ for
all $y\in\Z^d$ and $|\eta|\leq\eta^*$. Here $|C|=\sum_{\ga\in C}
|\ga|$ and $\La(C)=\cup_{\ga\in C} \La(\ga)$.
\end{lem}

\begin{rem}
It will turn out in the proof of this lemma (see
Appendix~\ref{sec: proof of cl.exp.}) that the dependence of the
constants $c_1,c_2$, and $\tau$ on $d,\la^*$, and $\eta^*$ is of
the form $\th\la^* + \eta^* + \text{const}(d)$. Similar
dependencies also occur in the next lemma.
\end{rem}


\begin{lem}\label{lem: cl.exp. +-}
There exist constants $\hat c_1,\hat c_2<\hat\tau \leq \tau$
depending on $d$ and $\eta^*$ such that for any $\be\geq\hat\tau$
and $x_0\in\Z^d$ one has
\begin{gather}\label{eq: hat phi-decay}
   \sum_{\hat C:\, \La(\hat C)\ni x}
      e^{2(\be-\hat c_1)|\hat C|}  |\Phi^{\pm,x_0}_{\hat C}|
      \leq 1 \\
   \itt{and} \label{eq: hat phi'-decay}
   \sum_{\hat C:\, \La(\hat C)\ni x}
      e^{2(\be-\hat c_2) |\hat C|} \Bigl|
      \frac{\dif\Phi^{\pm,x_0}_{\hat C}}{\dif\eta} \Bigr|
       \leq 1
\end{gather}
for all $x\in\La_n$ and $n\in\N$ whenever $|\eta|\leq\eta^*$. Here
$\tau$ is the constant from Lemma~\ref{Lem: cl.exp.}.
\end{lem}

In the following proposition we prove that the limits of
$\{\mu_{n}^{\la,\pm}\}$ and $\{\mu_{n}^{\pm}\}$ coincide on the
level of ``magnetizations''. Instead of proving that the limits
are actually identical, in Corollary~\ref{cor: equivalence} we use
an abstract argument to show that the limit points of
$\{\mu_{n}^{\la}\}$ coincide with $\mu^{\pm}$ whenever they
correspond to infinite limit points of $F_n^\la$.

\begin{prop}\label{prop: equivalence of measures}
Let $\be\geq\tau$, where $\tau$ is the constant from
Lemma~\ref{Lem: cl.exp.}. Then for every $x\in\Z^d$ one has
$\lim_{n\to\infty} \mu_n^{\la,\pm}(\si_x) = \mu^\pm(\si_x)$.
\end{prop}

\begin{proof}
Let $\be\geq \tau$ and $x\in\Z^d$. Taking into account the
convergent cluster expansions \eqref{eq: expansion Z la} and
\eqref{eq: expansion Z +-}, we have
\begin{gather}
  \mu_n^{\la,\pm} (\si_x) =
   \frac{\dif \log Z_n^{\la,x,\pm}}{\dif \eta} \Bigr|_{\eta=0}
   = \pm 1 +  \sum_{C: \, \La(C)\ni x}
   \frac{\dif \Phi_C^{\la,x,\pm}}{\dif \eta} \Bigr|_{\eta=0}  \\
\intertext{and}
  \mu_n^{\pm} (\si_x) =
   \frac{\dif \log Z_n^{\pm,x}}{\dif \eta} \Bigr|_{\eta=0}
   = \pm 1 + \sum_{\hat C: \, \La(\hat C)\ni x}
   \frac{\dif \Phi_{\hat C}^{\pm,x}}{\dif \eta} \Bigr|_{\eta=0}
\end{gather}
for all $n\geq n_x$.\footnote{Recall that $x\in\La_n$ for all
$n\geq n_x$.} Since any contour $\ga$ from $\cD_n$ whose volume
$\La(\ga)$ does not intersect $\dL_n$ is necessarily in $\hat
\cD_n$ and vice versa and since $K_\ga^{\la,x,\pm} =
K_\ga^{\pm,x}$ for such $\ga$, it follows that
\begin{equation}
   \mu_n^{\la,\pm} (\si_x) - \mu_n^\pm (\si_x) =
   \sum_{C: \, \La(C)\ni x \atop \La(C) \cap \dL_n \neq \emptyset}
   \frac{\dif \Phi_C^{\la,x,\pm}}{\dif \eta}\Bigr|_{\eta=0}
   - \sum_{\hat C: \, \La(\hat C)\ni x \atop \La(\hat C) \cap \dL_n
   \neq \emptyset} \frac{\dif \Phi_{\hat C}^{\pm,x}}{\dif \eta}
   \Bigr|_{\eta=0}
\end{equation}
for all $n\geq n_x$. Realizing that a cluster contributing to any
of the last two sums must necessarily satisfy $|C|\geq \frac n4$
whenever $n\geq 2n_x$, Lemma~\ref{Lem: cl.exp.} and \ref{lem:
cl.exp. +-} yield
\begin{multline}
  | \mu_n^{\la,\pm} (\si_x) - \mu_n^\pm (\si_x) | \leq
  e^{-(\be-\max\{c_2, \hat c_2\}) \frac n2} \; \times \\ \times
  \Bigl( \sum_{C: \La(C)\ni x} e^{2(\be-c_2)|C|}  \Bigl|\,
  \frac{\dif \Phi_C^{\la,x,\pm}}{\dif \eta} \, \Bigr| +
  \sum_{\hat C: \La(\hat C)\ni x} e^{2(\be-\hat c_2)|\hat C|}
  \Bigl|\, \frac{\dif \Phi_{\hat C}^{x,\pm}}{\dif \eta} \,
  \Bigr| \; \Bigr)_{\eta=0} \leq \\
  \leq 2 e^{-(\be-\max\{c_2, \hat c_2\}) \frac n2} .
\end{multline}
As a result, we have $\lim_{n\to\infty} | \mu_n^{\la,\pm} (\si_x)
- \mu_n^\pm (\si_x) | = 0$.
\end{proof}

\begin{cor}\label{cor: equivalence}
Let $\be \geq \tau$, where $\tau$ is the constant from
Lemma~\ref{Lem: cl.exp.}. If $k_n$ is an increasing sequence of
integers such that $\lim_n F_{k_n} = \pm\infty$, then
$\mu_{k_n}^{\la} \to \mu^{\pm}$ weakly.
\end{cor}

\begin{proof}
Let $\lim_n F_{k_n} = \infty$. Since $\lim_n
\mu_{k_n}^{\la,+}(\si_x) = \mu^{+}(\si_x)$ for all $x\in\Z^d$ due
to Proposition~\ref{prop: equivalence of measures}, it follows by
\eqref{eq: main relation} that also $\lim_n \mu_{k_n}^{\la}(\si_x)
= \mu^{+}(\si_x)$. Using a compactness argument, the sequence
$\mu_{k_n}^{\la}$ has a limit point in the weak topology. If
$\nu^{\la}$ is any such limit point, then it is FKG-dominated by
$\mu^{+}$ and satisfies $\nu^{\la}(\si_x) = \mu^{+}(\si_x)$ for
all $x$. As a consequence, see \cite{Li}, Corollary II.2.8, one
has $\nu^{\la} = \mu^{+}$, implying $\lim_n \mu_{k_n}^{\la} =
\mu^{+}$. The case $\lim_n F_{k_n} = -\infty$ then immediately
follows by the spin-flip symmetry.
\end{proof}

\section{Probabilistic Analysis}\label{sec: prob.anal.}

In view of Corollary~\ref{cor: equivalence}, the study of the
limit points of the sequence $\{\mu_n^\la\}$ boils down to the
analysis of the sequence of random functions $\{F_n^\la\}$. Using
\eqref{eq: expansion Z la} with $\eta = 0$, they have the form
\begin{equation}\label{eq: F_n}
   F_n^\la = 2S_n^\la + \sum_{C\in\dif\cC_n} \De \Phi_C^\la
\end{equation}
with $\dif\cC_n$ being the set of clusters from $\cC_n$ for which
$\cP(C)=\cup_{\ga\in C} \cP(\ga)$ is not empty. Notice that
$\De\Phi_C^\la$ only depends on $\la_x$ iff $x\in\dif_n(C)$.

In order to prove our main claim about the structure of the limit
points of the random sequence $\{F_n^{\la}\}$, see Proposition
\ref{prop: limit points of F}, we need a version of the
local-limit theorem for this sequence. If the actual local-limit
theorem would hold, we could conclude that there exists a sequence
of numbers $\{\al_n\}$, $\al_n = O(n^{d-1})$, such that
\begin{equation}\label{eq: would-like LLT}
  \lim_{n\to\infty} \al_n^{1/2} \Pro(F_n^{\la} \in (a,b)) = b - a
\end{equation}
for every finite $a < b$. If $F_n^\la$ were a sum of i.i.d.~random
variables (as is true in the toy model), such a strong result
could easily be derived \cite{Du}. However, the $\Delta
\Phi_{C}^{\lambda}$ terms in \eqref{eq: F_n} spoil the
independence, and we are not able to establish a statement of the
form \eqref{eq: would-like LLT}. Nevertheless, we can again apply
a cluster expansion, see \eqref{eq: log psi} below, now for the
boundary term, which will give us a slightly weaker result. More
precisely, we estimate from above the probabilities to find
$F_n^{\la}$ in intervals which are not fixed but rather grow as
small powers of $n$, see Lemma \ref{lem: weak LLT} below. This
weaker result is enough for what we need, which is that the free
energy differences between plus and minus due to the random
boundary term will be far enough away from zero for all large
enough volumes, with overwhelming probability. Therefore either
the plus or the minus state will dominate.

Let us consider the \emph{characteristic function}
\begin{equation}\label{eq: char fct}
   \psi_n (t) = \Exp \, e^{i t F_n^\la},  \quad t\in\R.
\end{equation}
In order to control $\psi_n(t)$ for small values of $t$, we
rewrite it as the partition function of a polymer model with
complex weights as follows. Realizing that
\begin{equation}
   \prod_{C\in\dif\cC_n} e^{it \De\Phi_C^\la} =
   \sum_{\mfr{C}\subset\dif\cC_n} \;
   \prod_{C\in\mfr{C}} \bigl( e^{it \De\Phi_C^\la} -1 \bigr),
\end{equation}
where the term corresponding to $\mfr{C}=\emptyset$ is set
equal to 1, and using $\dif_n(\mfr{C})$ to denote $\cup_{C\in\mfr C}
\dif_n(C)$, it readily follows that
\begin{equation}
   \psi_n (t) = \sum_{\mfr{C}\subset\dif\cC_n}
      \Bigl( \Exp\, e^{2it S_{\dif_n(\mfr{C})}^\la}
      \prod_{C\in\mfr C}
      \bigl( e^{it \De\Phi_C^\la} -1 \bigr) \Bigr) \;
      \Exp\, e^{2it S_{\dL_n\bsl \dif_n(\mfr C)}^\la}.
\end{equation}
Since $\Exp\, e^{2it S_{\dL_n\bsl \dif_n(\mfr C)}^\la} =
(\vp(2t))^{|\dL_n\bsl \dif_n(\mfr C)|}$, we therefore have
\begin{gather}\label{eq: psi as polymer - 1}
    \psi_n (t) = (\vp(2t))^{|\dL_n|}
    \sum_{\mfr C\subset\dif\cC_n} \vr_{\mfr C} (t)
    \itt{with} \label{eq: vr}
      \vr_{\mfr C} (t) =  (\vp(2t))^{-|\dif_n(\mfr C)|}\;
      \Exp\, e^{2it S_{\dif_n(\mfr{C})}^\la}
      \prod_{C\in\mfr C}
      \bigl( e^{it \De\Phi_C^\la } -1 \bigr)
\end{gather}
for all $t\in\R$ for which $\vp(2t)\neq 0$; we only consider such
$t$ in the sequel.

Let $\cG(\mfr{C})$ be the graph on the vertices of all clusters in
$\mfr{C}\subset\cC_n$ such that $C_1,C_2\in\mfr{C}$ are connected
by the edge iff $\cP(C_1)\cap\cP(C_2)\neq\emptyset$ (sharing of at
least one plaquette of $\dif V_n$). We say that $\mfr{C}$ is a
\emph{connected family of clusters} whenever the corresponding
graph $\cG(\mfr{C})$ is connected. Clearly,
\begin{equation}
   \vr_{\mfr C} (t) = \prod_{{\mfr C}_{\co} \subset\mfr C}
        \vr_{{\mfr C}_{\co}} (t) ,
\end{equation}
where the product runs over all connected components of the family
$\mfr{C}$. Writing $\cP(\mfr{C})=\cup_{C\in\mfr{C}} \cP(C)$ and
collecting all the connected families $\mfr{C}_{\co}$ of clusters
with the same set $\cP({\mfr C}_{\co})$, we get
\begin{multline}
     \sum_{\mfr C\subset\dif\cC_n} \vr_{\mfr C} (t) =
     \sum_{\{P_k\}} \; \sum_{{\mfr C}_{\co}^{(1)}:\,
           \cP({\mfr C}_{\co}^{(1)}) = P_1, \atop \vdots}
    \prod_k \; \vr_{{\mfr C}_{\co}^{(k)}} (t) = \\
    = \sum_{\{P_k\}} \; \prod_k \;
    \sum_{{\mfr C}_{\co}:\, \cP({\mfr C}_{\co}) = P_k}
    \vr_{{\mfr C}_{\co}} (t).
\end{multline}
In view of \eqref{eq: psi as polymer - 1}, we have thus rewritten
$\psi_n$ as the partition function of a polymer model,
\begin{equation}
    \psi_n (t) = (\vp(2t))^{|\dL_n|}
    \sum_{\{P_k\}}\; \prod_k \; w_{P_k} (t) ,
  \end{equation}
in which \emph{polymers} are any (not necessarily connected) sets
of plaquettes on $\dif V_n$, the polymer weights are
\begin{equation} \label{eq: polymer weights}
   w_P (t) = \sum_{{\mfr C}_{\co}:\, \cP({\mfr C}_{\co}) = P}
                \vr_{{\mfr C}_{\co}} (t),
\end{equation}
and \emph{incompatibility} of two distinct polymers means sharing
of at least one of their plaquettes. Denoting by $\frY_n$ the set
of all clusters of polymers in $V_n$, the characteristic function
is given by the cluster expansion
\begin{equation}\label{eq: log psi}
  \log \psi_n (t) = |\dL_n| \log\vp(2t) +
  \sum_{\mfr \cY \in \frY_n} w^T_{\cY} (t) \ .
\end{equation}

The following two statements establish a control over the
behaviour of the characteristic function in a neighbourhood of the
origin. A rather technical proof of Lemma~\ref{lem: char.fct.} is
given in Appendix~\ref{sec: proof of Lemma 5.1}.

\begin{lem}\label{lem: char.fct.}
There exist constants $\be_0 = \be_0(d,\la^*) < \infty$ and $\ep >
0$ such that $\be \geq \be_0$ implies the inequality
\begin{equation}\label{eq: bound on cl.}
  \sum_{\cY \in \frY_n} |w^T_{\cY} (t)| \leq
  \frac12 \, \si^2 t^2\, |\dif\La_n|
\end{equation}
for all $n$ and $|t|\leq\ep$. Here $\si^2 = \Exp \la_0^2$ is the
variance of the distribution of the boundary fields.
\end{lem}

\begin{cor}\label{cor: est.char.fct}
Let $\si^2 > 0$ and $\be \geq \be_0$ with $\be_0$ being the
constant from Lemma~\ref{lem: char.fct.}. Then there exists $t_0 >
0$ such that
\begin{equation}
   |\psi_n(t)| \leq \exp \Bigl(-\frac12 \, \si^2 t^2 |\dL_n| \Bigr)
\end{equation}
holds true for any $n$ and $|t| \leq t_0$.
\end{cor}

\begin{proof}
Since $\log \vp(t) = -\frac{1}{2} \si^2 t^2 + o(t^2)$ and since
$\si^2 > 0$, there exists $t_1 > 0$ such that $|\log \vp(t) +
\frac{1}{2} \si^2 t^2| \leq \frac{1}{4} \si^2 t^2$ whenever $|t|
\leq t_1$, yielding
\begin{equation}
  |\vp(t)| \leq e^{-\frac{1}{2}\si^2 t^2}\,
  e^{|\log \vp(t) + \frac{1}{2}\si^2 t^2|} \leq
  e^{-\frac{1}{4}\si^2 t^2}\ .
\end{equation}
Using the cluster expansion \eqref{eq: log psi} and
Lemma~\ref{lem: char.fct.}, we immediately get the above statement
with $t_0 = \min \{t_1,\ep\}$.
\end{proof}

We now prove the following weak variant of the local-limit
theorem.

\begin{lem}\label{lem: weak LLT}
Let the assumptions of Corollary~\ref{cor: est.char.fct} be
satisfied. Then, for any finite interval $\cI \subset \R$ whose
end-points are $a$ and $b$ and any $\ze > 0$, we have
\begin{equation}\label{eq: estimate}
  \limsup_{n \to \infty} \, n^{\frac{d-1}{2} - \ze} \; \Pro \left(
  F_{n}^{\lambda} \in n^{\ze} \cI \right) < \infty\ .
\end{equation}
Here $n^\ze \cI$ is the interval with the end-points $a n^\ze$ and
$b n^\ze$.
\end{lem}

\begin{proof}
The idea of the proof is to ``blur'' the distribution function of
the random variable $F_{n}^{\lambda}$ by convoluting it with a
smooth function without changing the inequality \eqref{eq:
estimate}. This trick will enable us to obtain a sufficient
control over the asymptotic behaviour of the characteristic
function outside the regime where the cluster expansions hold.

Let $g \in C^{\infty}$ be a positive function with a compact
support in $[-1,1]$ and satisfying the normalization condition
$\int_{\R} g(x) \, dx = 1$. Further, we use $\Dis_n$ to denote the
distribution function of $F_{n}^{\la}$ and define
\begin{equation}\label{eq: blur}
  \tilde{\Dis}_n(z) = \int_{-\infty}^{z} dx
  \int_{\R} g_{n}(x-y) \, d\Dis_n(y)\ ,
\end{equation}
 where
\begin{equation}
  g_{n}(x) = n^{-\ze} g(xn^{-\ze})
\end{equation}
with $\ze > 0$. The function $\tilde{\Dis}_n$ is clearly a
distribution function due to the properties of $g$. Given an
interval $\cI\subset\R$ with the end-points $a \leq b$, the lemma
will be proved once we show that
\begin{equation}\label{eq: modified}
  \limsup_{n \to \infty} \, n^{\frac{d-1}{2} - \ze}
  \int_{n^{\ze} \tilde\cI} d\tilde{\Dis}_n < \infty ,
\end{equation}
where $\tilde{\cI} = [a-1,b+1]$. Indeed, since
\begin{equation}\label{eq: estim.above}
\begin{split}
  \int_{n^\ze \tilde\cI} d\tilde{\Dis}_n &=
  \int_{n^\ze \tilde\cI} dx \int_{\R}
  g_n(x-y) \, d\Dis_n(y) \geq
\\
  &\geq \int_{n^\ze \tilde\cI} dx \int_{n^{\ze} \cI}
  g_n(x-y) \, d\Dis_n(y) =
\\
  &= \int_{n^{\ze} \cI} d\Dis_n(y) \int_{\R} g_n(x-y) \, dx =
  \int_{n^{\ze} \cI} d\Dis_n
\end{split}
\end{equation}
by Fubini's theorem and the normalization condition $\int_{\R}
g_n(x) dx = 1$, this estimate combined with \eqref{eq:
estim.above} immediately yields the lemma.

Turning now to the proof of \eqref{eq: modified}, we first
introduce the functions
\begin{equation}
  \hat{g}(t) = \int_{\R} e^{ixt} g(x) \, dx
\end{equation}
and $\hat{g}_{n}(t) = \hat{g}(tn^{\ze})$. Moreover, since $g \in
C^{\infty}$ and has a compact support, for all $k = 0,1,2,\ldots$
and $t \in \R\setminus\{0\}$ the bounds $|\hat{g}(t)| \leq c_k
|t|^{-k}$ are true, where $c_0 = 1$ and $c_k < \infty$ for $k \geq
1$, implying
\begin{equation}\label{eq: g-est.}
  |\hat{g}_{n}(t)| \leq c_k n^{-k\ze} |t|^{-k}
\end{equation}
for all $n\in\N$. It also immediately follows that the modified
characteristic function
\begin{equation}\label{eq: mod.char.fct.}
  \tilde{\psi}_{n}(t) = \int_{\R} e^{ixt} \, d\tilde{\Dis}_n(x) =
  \psi_n(t) \hat{g}_{n}(t)
\end{equation}
satisfies the condition $\int_{\R} |\tilde{\psi}_n(t)| \, dx <
\infty$ and, therefore, $\tilde{\Dis}_n$ is given by the inversion
formula \begin{equation}
  \tilde{\Dis}_n(z) = \int_{-\infty}^{z} \frac{dx}{2\pi}
  \int_{\R} e^{-itx} \tilde{\psi}_n(t) dt\ .
\end{equation}
Using \eqref{eq: mod.char.fct.} and the fact that $|\psi_n(t)|
\leq 1$, we obtain the estimate
\begin{equation}
\begin{split}
  n^{\frac{d-1}{2} - \ze} &\int_{\tilde{\cI}n^{\ze}} d\tilde{\Dis}_n =
  n^{\frac{d-1}{2} - \ze} \int_{\tilde{\cI}n^{\ze}} \frac{dx}{2\pi}
  \int_{\R} e^{-itx} \, \tilde{\psi}_n(t) \, dt \leq
\\
  &\leq n^{\frac{d-1}{2}} \int_{\tilde{\cI}} \frac{dx}{2\pi} \int_{\R}
  |\tilde{\psi}_n(t)| \, dt \leq
  (I_n^1 + I_n^2) \int_{\tilde{\cI}} \frac{dx}{2\pi}
\end{split}
\end{equation}
for all $n\in\N$. Here
\begin{equation}
  I_n^1 = n^{\frac{d-1}{2}} \int_{|t| \leq t_0} |\psi_n(t)| \, dt
  \quad \text{and} \quad
  I_n^2 = n^{\frac{d-1}{2}} \int_{|t| > t_0} |\hat{g}_{n}(t)| \, dt
\end{equation}
with $t_0$ being the constant from Corollary~\ref{cor:
est.char.fct}. The interval $\tilde\cI$ being finite, it now
suffices to show that the integrals $I_n^1$ and $I_n^2$ are
uniformly bounded if $n\to\infty$. First, in view of \eqref{eq:
g-est.}, one can conclude that
\begin{equation}
  I_n^2  \leq  c_k n^{\frac{d-1}{2} - k\ze}
  \int_{|t| > t_0} |t|^{-k} \, dt .
\end{equation}
Hence, choosing an integer $k > \max\{1,\frac{d-1}{2\ze}\}$, we
get $\lim_{n\to\infty} I_n^2 = 0$. In order to estimate the
integral $I_n^1$, we make use of Corollary~\ref{cor: est.char.fct}
to obtain
\begin{equation}
  \limsup_{n\to\infty} I_n^1  \leq
  \limsup_{n\to\infty} n^{\frac{d-1}{2}} \int_{\R}
  \exp \Bigl( -\frac12\, \si^2 t^2 |\dL_n| \Bigr) \, dt
  = \frac1\si \, \Bigl( \frac\pi d \Bigr)^{1/2} ,
\end{equation}
which finishes the proof.
\end{proof}

The proof of Theorem~\ref{thm: main} will be finished once we
prove the following lemma, yielding the structure of limit points
of the sequence $F_n^{\la}$. For convenience, we use $\frL^{\la}$
to denote the (random) set of all limit points of the sequence
$\{F_{n}^{\la}\}$ and $\frL^{\la}_{d,\,\om}$ for the set of limit
points of the ``sparse'' sequence
$\{F_{n^{[4-d+\omega]}}^{\la}\}$.

\begin{prop}\label{prop: limit points of F}
Let $\si^2 > 0$ and $\be \geq \be_0$ with $\be_0$ from
Lemma~\ref{lem: char.fct.}.
\begin{enumerate}
\item
  If $d > 3$, then $\frL^{\la} = \{\infty,-\infty\}$ $\Pro$-a.s.
\item
  If $d \in \{2,3\}$ and $\omega > 0$ , then
  $\frL^{\la}_{d,\,\om} = \{\infty,-\infty\}$ $\Pro$-a.s.
\end{enumerate}
\end{prop}

\begin{proof}
(1) Let $d > 3$. First, we shall show that $\frL^{\la} \cap \R =
\emptyset$ a.s. Defining the events
\begin{equation} \label{eq: events E}
  \cE_{n,k}^{\ze} =
  \{\la: \, -k n^{\ze} < F_{n}^{\la} < k n^{\ze} \}
\end{equation}
for all $k\in\N$ and $\ze\geq 0$, Lemma~\ref{lem: weak LLT}
implies that there are constants $c_k (\ze),n_k (\ze)<\infty$ such
that
\begin{equation}
  \Pro(\cE_{n,k}^{\ze}) \leq c_k (\ze) \, n^{-\frac{d-1}{2} + \ze}
\end{equation}
for any $\ze > 0$ whenever $n \geq n_k(\ze)$. Choosing
$0 < \ze < \frac{d-3}{2}$, this yields
\begin{equation}\label{eq: summable}
  \sum_n \Pro(\cE_{n,k}^{0}) \leq \sum_n \Pro(\cE_{n,k}^{\ze}) < \infty\ .
\end{equation}
Using the Borel-Cantelli lemma, it follows that $\Pro(\limsup_n
\cE^{0}_{n,k}) = 0$, where $\limsup_n \cE^{0}_{n,k} = \cap_n
\cup_{m=n}^{\infty} \cE_{m,k}^{0}$ is the event that infinitely
many events $\cE_{n,k}^{0}$ occur. As a consequence, we have
$\Pro(\frL^{\la} \cap (-k,k) \neq \emptyset) = 0$. Hence,
\begin{equation}\label{eq: no limit}
  \Pro(\frL^{\la} \cap \R \neq \emptyset) \leq
  \sum_{k} \Pro(\frL^{\la} \cap (-k,k) \neq \emptyset) = 0\ .
\end{equation}

Further, the events $\cA_n^{+} = \{\la:\, F_n^{\la} \geq 0\}$ and
$\cA_n^{-} = \{\la:\, F_n^{\la} \leq 0\}$ satisfy
$\lim_{n\to\infty} \Pro(\cA_{n}^{\pm}) = \frac{1}{2}$ due to the
symmetry of the distribution $\Pro$ and because $\lim_{n\to\infty}
\Pro(\la: F^\la_n = 0) = 0$ by the the same argument as above.
Since $\{\cA_{2n}^\pm \}$ are subsequences of independent events,
one gets $\Pro(\limsup_n \cA_n^{\pm}) = 1$ by the (second)
Borel-Cantelli lemma. Therefore, both $F_n^\la \geq 0$ and
$F_n^\la \leq 0$ occur infinitely many times $\Pro$-almost surely
and we get $\Pro(\frL^\la \cap [0,\infty] \neq\emptyset) = 1$ as
well as $\Pro(\frL^\la \cap [-\infty,0] \neq\emptyset) = 1$.
Combined with \eqref{eq: no limit}, this proves the statement.

(2) Let $d \in \{2,3\}$. Recalling the definition \eqref{eq:
events E} of the events $\cE_{n,k}^{\ze}$, this time one arrives
at the inequality
\begin{equation}
  \sum_n \Pro(\cE_{[n^{4-d+\om}],\,k}^{0}) < \infty
\end{equation}
whenever choosing $0<\ze<\frac{d-1}2-\frac1{4-d+\om}$. The rest
of the proof runs along the same lines as for $d>3$.
\end{proof}

\begin{cor}\label{cor: metastates}
In the situation of Proposition 5.4, the Newman-Stein metastate is
${{1 \over 2} (\delta_{\mu^{+}} + \delta_{\mu^{-}})}$.
\end{cor}

\begin{rem}
In fact we have proved something strictly stronger. If the set of
mixed states is null-recurrent (which we suspect happens for
non-sparse sequences in $d = 2,3$), this still would lead to the
same metastate. However, this we cannot prove.
\end{rem}

\section{Concluding Remarks}\label{sec: concl. remarks}

In the introduction, we have concluded that the set of limit
points of $\{\mu^{\la}_{n}\}$ for $d = 2,3$ and $\be=\infty$ is
countable a.s., containing all the convex combinations
$\frac{1+\al}2\, \de_1 + \frac{1-\al}2\, \de_{-1}$ whenever $\al=
\tanh \al'$ for some $\al'\in 2\Z \cup \{-\infty,\infty\}$.
Nevertheless, once $\be_0 \leq \be < \infty$, our conjecture is
that any convex combination of $\mu^+$ and $\mu^-$ is a limit
point of $\{\mu^{\la}_{n}\}$ a.s. In order to verify this, one
would need to show that
$$
   \liminf_{n\to\infty} \; n^{\frac{d-1}{2}+\ze} \,
     \Pro(F_n^\la \in (a,b)) > 0
$$
for a sufficiently small $\ze\geq 0$ whenever $a<b$. In the
context of our perturbation scheme, this would require a variant
of Lemma~\ref{lem: cl.exp.-rozdiely}, yielding a lower bound on
the cluster-weight differences $\De\Phi_C^\la$. The problem of
lower bounds on cluster weights is highly non-trivial, however.

Presumably, Theorem~\ref{thm: main} remains valid for all
symmetric distributions with zero mean and a positive variance,
provided $\be$ is large enough (depending only on $d$). It is the
uniformity of cluster expansions in realizations of the boundary
fields $\la$ why we restrict ourselves to $\la$ of small strength
in the paper. As a result, one has an extra attraction of contours
to the boundary (suppressing interfaces even for the Dobrushin
boundary field realizations) which always leads to a convex
combination of the two translation-invariant extremal states. In
order to prove the theorem for all distributions of $\la$, one
should perhaps refine the strategy by replacing the uniformity
with the ``typicality'' and by using a coarse-graining argument to
show that large-scale contours are typically suppressed or
attracted to the boundary even without the above-mentioned extra
attraction. We defer the details to a later investigation.

It is interesting to consider the situation of asymmetric
distributions of $\la$ (keeping the mean zero). Clearly, whenever
$\liminf_n \frac{1}{|\dif\La_n|}\, \Exp\, F_n^\la > 0$ or
$\limsup_n \frac{1}{|\dif\La_n|} \, \Exp\, F_n^\la < 0$, then
$\mu_n\to\mu^+$ a.s.\ or $\mu_n\to\mu^-$ a.s., respectively. By
adding a boundary ``magnetic-field'' term to the Hamiltonian, one
can find a transition between these two regimes. Heuristically,
taking into account only the clusters surrounding a single site,
the leading asymptotics of the transition point is $h \approx
e^{-2(2d-1)\be} \, \Exp\, \sinh \la_0$. Yet, this time one cannot
conclude whether the chaotic size-dependence actually occurs at
the transition point because Lemma~\ref{prop: limit points of F},
where the symmetry of the distribution plays a crucial role, is
not valid any more.

\appendix
\section{Cluster expansions}\label{sec: clust.exp.}

In this appendix we summarize statements on the convergence of
cluster expansions for the abstract polymer model in the context
of the Koteck\'y-Preiss formalism \cite{KoPr}. Besides the
standard result concerning the exponential decay of cluster
weights, we consider a model with parameter-dependent weights
and prove estimates on their derivatives. The proof proposed here
is entirely in the spirit of \cite{KoPr}.

Let $\cK$ be a countable set and let us call its elements
\emph{polymers}. Given a reflexive and symmetric relation $\io
\subset \cK\times\cK$, two polymers $\Ga_1,\Ga_2\in\cK$ are
\emph{incompatible} if $\Ga_1 \io \Ga_2$; otherwise they are
\emph{compatible}. A finite set $\De \subset \cK$ is called
compatible whenever all polymers from $\De$ are pairwise
compatible. If $\De$ is not a union of two disjoint, non-empty
sets $\De_1$ and $\De_2$ such that $\Ga_1$ is compatible with
$\Ga_2$ for any $\Ga_1\in\De_1$ and $\Ga_2\in\De_2$, then $\De$ is
a \emph{cluster}. We write $\De\io\Ga$ whenever there exists
$\Ga'\in\De$ such that $\Ga'\io\Ga$.

Let us consider a function $w:\cK \to \C$ called \emph{weight}.
For any finite set $\De\subset\cK$, we let
\begin{equation}
   w(\De) = \begin{cases}
          \prod_{\Ga\in\De} w(\Ga) & \text{if $\De$ is compatible}, \\
          0 & \text{otherwise.}
   \end{cases}
\end{equation}
The \emph{partition function} in a finite set $\La\subset\cK$ is
defined as
\begin{equation}
   \cZ (\La) = \sum_{\De\subset\La} w(\De) \ .
\end{equation}
Its logarithm can be formally written in the form
\begin{gather}\label{eq: log Z}
   \log \cZ (\La) = \sum_{\De\subset\La} w^T (\De) \ ,  \\
\intertext{where the weights $w^T$ are unique and given by the
M\"obius inversion formula}\label{eq: wT}
   w^T (\De) = \sum_{\La \subset \De} (-1)^{|\De \setminus \La|}
                       \log \cZ (\La)
\end{gather}
for any finite $\De\subset\cK$. As a consequence, one has $w^T
(\De)=0$ whenever $\De$ is not a cluster.

\begin{prop} \label{prop: cl.exp.}
Given functions $a,b:\cK \to [0,\infty)$, let the condition
\begin{equation}\label{eq: KP-general}
   \sum_{\Ga \io \Ga_0} e^{(a+b)(\Ga)} |w(\Ga)| \leq a(\Ga_0)
\end{equation}
be satisfied for every $\Ga_0 \in \cK$. Then
\begin{equation} \label{eq: KP}
   \sum_{\De \io \Ga_0} e^{b(\De)} |w^T (\De)| \leq a(\Ga_0)
\end{equation}
with $b(\De)=\sum_{\Ga\in\De} b(\Ga)$. Moreover, let the weights
$w$ be differentiable functions in an open interval $\cI \subset
\R$. If the condition \eqref{eq: KP-general} is true uniformly in
$\cI$ with $a \leq b$ and if $c: \cK \to [0,\infty]$ is a function
such that\footnote{
     We allow $c(\Ga)$ to be $\infty$ and use the convention
     $0 \cdot \infty = 0$ in order to cover the contour models
     introduced in Section~\ref{sec: contour rep.}.}
\begin{equation}\label{eq: KPpodm-der}
   \sum_{\Ga \io \Ga_0} c(\Ga)\,e^{(a+b)(\Ga)} \Bigl|
            \frac{dw(\Ga)}{d\eta} \Bigr| \leq a(\Ga_0) \ ,
\end{equation}
then
\begin{equation}\label{eq: KP'}
  \sum_{\De\io\Ga_0} c(\De)\,e^{(b-a)(\De)} \Bigl|
     \frac{d w^T (\De;\eta)}{d\eta}  \Bigr|
     \leq 2a(\Ga_0)
\end{equation}
for any $\Ga_0 \in \cK$ and $\eta \in \cI$. Here, $c(\De) =
\min_{\Ga\in\De} c(\Ga)$.
\end{prop}

\begin{proof}
The bound \eqref{eq: KP} is proved in \cite{KoPr}. The proof of
\eqref{eq: KP'} goes along the same lines as follows.

Let $\Ga_0\in\cK$ be fixed. Let $w_s(\Ga)$ be the $s$-dependent
weight defined by $sw(\Ga)$ if $\Ga\io\Ga_0$ and by $w(\Ga)$
otherwise. Using the M\"obius formula \eqref{eq: wT}, we obtain
\begin{equation}
\begin{split}
  \frac{d w^T_s(\De;\eta)}{d\eta} &= \sum_{\La\subset\De}
  (-1)^{|\De\setminus\La|} \sum_{\Ga\in\La}
  \frac{d \log \cZ_s (\La)}{d w_s(\Ga)} \, \frac{d w_s(\Ga;\eta)}{d\eta} = \\
  &= \sum_{\Ga\in\De} \frac{d w_s(\Ga;\eta)}{d\eta}
     \sum_{\La:\, \Ga\in\La\subset\De} (-1)^{|\De\setminus\La|}
     \frac{\cZ_s(\La \setminus [\Ga])}{\cZ_s(\La)} .
\end{split}
\end{equation}
Here $[\Ga]$ is the set of all polymers incompatible with $\Ga$
and, due to the formula \eqref{eq: log Z}, we have the cluster
expansion
\begin{equation}\label{eq: aux1}
 \frac{\cZ_s (\La \setminus [\Ga])}{\cZ_s (\La)}  =
     \exp \Bigl[ - \sum_{\De'\subset\La \atop \De' \io\Ga} w^T_s (\De')
     \Bigr] .
\end{equation}
Following the strategy of \cite{KoPr}, we write
\begin{equation}\label{eq: aux2}
  \frac{d}{ds} \sum_{\De\io\Ga_0} c(\De)\,e^{(b-a)(\De)} \Bigl|
  \frac{d w^T_s (\De)}{d\eta} \Bigr| =  X_1 + X_2 \ ,
\end{equation}
where, in view of \eqref{eq: aux1} and \eqref{eq: aux2},
\begin{multline}
   X_1 = \sum_{\De\io\Ga_0} \ve(\De) c(\De)\,e^{(b-a)(\De)}
      \sum_{\Ga\in\De\atop\Ga\io\Ga_0}
      \frac{d w(\Ga)}{d\eta} \times
\\
      \times \sum_{\La:\, \Ga\in\La\subset\De}
      (-1)^{|\De\setminus\La|} \exp \Bigl[
      - \sum_{\De'\subset\La \atop \De' \io\Ga} w^T_s (\De') \Bigr] =
\\
      = \sum_{\Ga\io\Ga_0} \frac{d w(\Ga)}{d\eta}
      \sum_{\De\ni\Ga} \ve(\De) c(\De)\,e^{(b-a)(\De)}
      \sum_{n=0}^\infty \frac1{n!}
      \sum_{\De_1,\dots,\De_n \subset\De \atop
      \De_k \io\Ga,\; k=1,\dots,n}  \prod_{k=1}^n
      \bigl[ - w^T_s (\De_k) \bigr] \times
\\    \times \sum_{\La:\, \{\Ga\}\cup \bigcup_k \De_k \subset\La\subset\De}
      (-1)^{|\De\setminus\La|}  \ ,
\end{multline}
and
\begin{multline}
  X_2 = - \sum_{\De\io\Ga_0} \ve(\De) c(\De)\,e^{(b-a)(\De)}
       \sum_{\Ga\in\De}  \frac{d w_s(\Ga)}{d\eta} \times \\ \times
       \sum_{\La:\, \Ga\in\La\subset\De} (-1)^{|\De\setminus\La|}
       \sum_{\De'\subset\La \atop \De'\io\Ga,\,\Ga_0}
       \frac{ d w^T_s (\De')}{ds} \exp \Bigl[
      - \sum_{\De''\subset\La \atop \De'' \io\Ga} w^T_s (\De'') \Bigr] =
\\
      = \sum_{\De'\io\Ga_0} \frac{d w^T_s (\De')}{ds}
      \sum_{\Ga\io\De'} \frac{d w_s(\Ga)}{d\eta}
      \sum_{\De\ni\Ga \atop \De\supset\De'} \ve(\De) c(\De)\,
      e^{(b-a)(\De)} \times
\\
      \times \sum_{n=0}^\infty \frac1{n!}
      \sum_{\De_1,\dots,\De_n \subset\De \atop \De_k \io\Ga,\,
              k=1,\dots,n}  \prod_{k=1}^n \bigl[ - w^T_s (\De_k) \bigr]
      \sum_{\La:\, \De'\cup \{\Ga\} \cup \bigcup_k \De_k \subset\La\subset\De}
      (-1)^{|\De\setminus\La|}  \ .
\end{multline}
Here $\ve(\De)$ is such that $\bigl| \frac{d w^T (\De)}{d\eta}
\bigr| = \ve(\De) \, \frac{d w^T (\De)}{d\eta}$. Since
$\sum_{B\subset A} (-1)^{|A \setminus B|}$ equals $1$ if
$A=\emptyset$ and $0$ otherwise, one can use the obvious estimates
$b(\cup_k A_k) \leq \sum_k b(A_k)$ and $c(\De) \leq c(\Ga)$ for
any $\Ga \in \De$ to get
\begin{multline}
  |X_1| \leq \sum_{\Ga\io\Ga_0}
     c(\Ga)\,e^{(b-a)(\Ga)} \Bigl| \frac{d w(\Ga)}{d\eta} \Bigr|
     \exp \Bigl[ \sum_{\De'\io\Ga} e^{(b-a)(\De')} |w^T_s(\De')|
     \Bigr] \leq
\\
     \leq \sum_{\Ga\io\Ga_0} c(\Ga)\,e^{b(\Ga)} \,
     \Bigl| \frac{d w(\Ga)}{d\eta} \Bigr| \leq a(\Ga_0)
\end{multline}
for any $s \in (0,1)$ due to \eqref{eq: KP} and \eqref{eq:
KPpodm-der}. Similarly,
\begin{multline}
  |X_2| \leq \sum_{\De'\io\Ga_0} e^{(b-a)(\De')}
     \Bigl| \frac{d w^T_s (\De')}{ds}
     \Bigr| \sum_{\Ga\io\De'} \Bigl|
     \frac{d w_s(\Ga)}{d\eta} \Bigr| \times
\\
     \times  \exp \Bigl[ \sum_{\De'\io\Ga}
      c(\Ga)\, e^{(b-a)(\De')} |w^T_s(\De')| \Bigr] \leq
\\
     \leq \sum_{\De'\io\Ga_0} e^{(b-a)(\De')} \Bigl|
     \frac{d w^T_s (\De')}{ds} \Bigr|
     \sum_{\Ga'\in\De'} \sum_{\Ga\io\Ga'} c(\Ga)\,e^{a(\Ga)}
     \Bigl| \frac{d w_s(\Ga)}{d\eta} \Bigr| \leq
\\
     \leq \sum_{\De\io\Ga_0} a(\De) e^{(b-a)(\De)}
     \Bigl| \frac{d w^T_s(\De)}{ds} \Bigr|.
\end{multline}
Using the inequality
\begin{equation}
  \frac{d}{ds} \sum_{\De\io\Ga_0} e^{b(\De)} |w^T_s (\De)| \leq a(\Ga_0)
\end{equation}
for any $s \in (0,1)$ proved in \cite{KoPr}, we may conclude that
$|X_2|\leq a(\Ga_0)$. To finish the proof, it now suffices to
realize that
\begin{equation}
  \sum_{\De\io\Ga_0} e^{(b-a)(\De)} \Bigl| \frac{dw^T(\De)}{d\eta}
         \Bigr| \leq
  \sup_{s\in(0,1)} \frac{d}{ds} \sum_{\De\io\Ga_0} e^{(b-a)(\De)}
    \Bigl| \frac{dw^T_s (\De)}{d\eta} \Bigr|  \ .
\end{equation}
\end{proof}

As an application of the above proposition, let us consider a
couple of weight functions $w_{1,2}:\cK\to\C$. Then the above
proposition implies an estimate on the difference of the
corresponding cluster weights $w^T_1$ and $w^T_2$.

\begin{cor}\label{cor: rozdiely}
Let $a,b : \cK\to[0,\infty)$, $a\leq b$, and let the condition
\eqref{eq: KP-general} be satisfied for both polymer weights $w_{1,2}$.
If there is a function $c: \cK \to [0,\infty]$ such that
\begin{equation}
  \sum_{\Ga\io\Ga_0} c(\Ga)\, e^{(a+b)(\Ga)}
  |(w_2 - w_1)(\Ga)| \leq a(\Ga_0)
\end{equation}
is true for all $\Ga_0\in \cK$, then
\begin{equation}\label{eq: KP''}
   \sum_{\De\io\Ga_0} c(\De)\, e^{(b-a)(\De)}
   |(w^T_2 - w^T_1)(\De)| \leq 2a(\Ga_0)\ .
\end{equation}
\end{cor}

\begin{proof}
The parameter-dependent weight $w(\eta)=\eta w_2 + (1-\eta) w_1$
satisfies the conditions \eqref{eq: KP} and \eqref{eq: KPpodm-der}
uniformly in the interval $[0,1]$. Therefore, using the inequality
\begin{equation}
  \sum_{\De\io\Ga_0} c(\De)\,e^{(b-a)(\De)}  |( w^T_2 - w^T_1)(\De)|
    \leq \sup_{\eta\in(0,1)} \sum_{\De\io\Ga_0} c(\De)\,e^{(b-a)(\De)}
    \Bigl| \frac{d w^T (\eta)}{d\eta} \Bigr|
\end{equation}
and \eqref{eq: KP'}, one immediately obtains \eqref{eq: KP''}.
\end{proof}

\section{Proof of Lemmas~\ref{Lem: cl.exp.} and
\ref{lem: cl.exp. +-}} \label{sec: proof of cl.exp.}

We begin with a geometrical lemma giving an estimate on the size
of the set $\cP(\ga)$ of all plaquettes of the boundary $\dif\,
\ol{\Inn\ga}$ lying on $\dif V_n$.


\begin{lem}\label{Lem: geometrical}
Let $d\geq 2$ and $n\in\N$. For an arbitrary contour $\ga\in\cD_n$
the estimate $|\cP(\ga)| \leq \th \, |\ga|$ with $\th=
\frac{2^{1/d}+1}{2^{1/d}-1}$
holds true.
\end{lem}

\begin{proof}
See Lemma B.3 in \cite{BK95}.
\end{proof}
\medz

\begin{proof}[Proof of Lemma~\ref{Lem: cl.exp.}.]
Let $\be\geq\tau$ with $\tau$ to be specified later. Let
$x_0\in\Z^d$ and $x\in\La_n$ be given. Together with the
inequalities \eqref{eq: phi-decay} and \eqref{eq: phi'-decay}, we
simultaneously prove that
\begin{gather}  \label{eq: K-decay}
   \sum_{\ga:\, \La(\ga)\ni x} e^{2(\be-c'_1) |\ga|}
         K^{\la,x,\pm}_{\ga} \leq 1 \\
   \itt{and} \label{eq: K'-decay}
   \sum_{\ga:\, \La(\ga)\ni x} e^{2(\be-c'_2) |\ga|} \Bigl|
         \frac{\dif K^{\la,x,\pm}_{\ga}}{\dif \eta} \Bigr|
       \leq 1
\end{gather}
for some constants $c'_1, c'_2 < \tau$ (depending on $d,\la^*$,
$\eta^*$). We shall proceed by induction on the size of the
volumes $v(\ga)$ and $v(C)=\max_{\ga\in C} v(\ga)$.

First, let us consider only contours and clusters with $v(\ga)=0$
and $v(C)=0$, respectively. From \eqref{eq: Z,K for 1-contour} it
follows that
\begin{equation}
  K_\ga^{\la,x_0,\pm}
  \leq  e^{-2 \left( \be|\ga| - |S_{\La(\ga)}^\la| - \eta^* \right)}
  \leq  e^{-2(\be - \th\la^* - \eta^* )|\ga|} ,
\end{equation}
where we also used Lemma~\ref{Lem: geometrical}. Since $|\ga|\geq
d$ and since there exists a constant $\vk=\vk(d)<\infty$ such that
the number of all contours $\ga\in\cD_n$ with $|\ga|=\ell$ and
such that $\La(\ga)$ contains a given site from $\La_n$ can be
bounded by $\vk^\ell$, the last estimate implies
\begin{equation}
  \sum_{\ga:\, \La(\ga)\ni x \atop v(\ga)=0} e^{2(\be-c'_1) |\ga|}
       K^{\la,x_0,\pm}_\ga
  \leq \sum_{\ell=d}^\infty
       (\vk e^{-2(c'_1 - \th\la^* - \eta^*)})^\ell
  \leq \frac12,
\end{equation}
provided $c'_1 - \th\la^* - \eta^* \geq \frac12\, \log 2\vk$, say,
which in its turn yields
\begin{equation}
  \sum_{\ga' \io \ga: \atop v(\ga')=0}
  e^{2(\be-c'_1)|\ga'|} K_{\ga'}^{\la,x_0,\pm}
  \leq |\ga| \, \max_{p\subset V_n} |\{x\in\La_n:\,
     \Box_x \cap p \neq \emptyset \}| \leq 3^d |\ga|.
\end{equation}
The condition \eqref{eq: KP-general} is thus satisfied in our case
with $a(\ga)= |\ga|$ and $b(\ga)=[2(\be - c'_1)-d\log 3
-1]\,|\ga|$. Hence, in view of Proposition~\ref{prop: cl.exp.}, we
have
\begin{multline}\label{eq: aux-ref}
  \sum_{C:\, \La(C)\ni x \atop v(C)=0} e^{2(\be-c_1)|C|}
     |\Phi_C^{\la,x_0,\pm}| \leq
  \sum_{\ga:\, \La(\ga)\ni x \atop v(\ga)=0} \;
  \sum_{C:\, C \ni \ga \atop v(C)=0}
     e^{2(\be-c_1)|C|}  |\Phi_C^{\la,x_0,\pm}| \leq \\
  \leq \sum_{\ga:\, \La(\ga)\ni x \atop v(\ga)=0}
     e^{-[2(c_1-c'_1)-d\log 3-1]\,|\ga|}
  \sum_{C:\, C \io \ga \atop v(C)=0} e^{[2(\be-c'_1)-d\log 3-1]\,|C|}
     |\Phi_C^{\la,x_0,\pm}| \leq \\
  \leq \sum_{\ga:\, \La(\ga)\ni x \atop v(\ga)=0}
     e^{-[2(c_1-c'_1)-d\log 3-2]\,|\ga|}
     \leq  2 (3^d\vk e^{-2(c_1-c'_1-1)})^d \leq 1
\end{multline}
if $c_1-c'_1-1\geq \frac12\, \log(2.3^d\vk)$. By virtue of
\eqref{eq: Z,K for 1-contour}, one has $\bigl|\, \frac{\dif}{\dif
\eta}\, K_\ga^{\la,x_0,\pm} \,\Bigr| \leq 2K_\ga^{\la,x_0,\pm}$.
Combined with the above, we therefore verified the inequalities
\eqref{eq: phi-decay} to \eqref{eq: K'-decay} for the considered
contours and clusters, providing that $c'_1 \geq \th\la^* + \eta^*
+ \frac12\, \log 2\vk$, $c_1 \geq c'_1+1+ \frac12\, \log(
2.3^d\vk)$, $c'_2 \geq \th\la^* + \eta^* + \frac12\, \log 2\vk$,
and $c_2 \geq c'_2+1+ \frac12\, \log( 2.3^d\vk)$.

Next, let us prove these inequalities for any contours and
clusters with $v(\ga)=N$ and $v(C)=N$, respectively, assuming that
they have already been proved for all contours and clusters with
their volumes smaller than $N$. Recalling that for any
$\ga'\subset\Inn\ga$ one necessarily has $v(\ga')<v(\ga)$, from
the inductive assumption it follows that $\log
Z_\ga^{\la,x_0,\pm}$ can be controlled by convergent cluster
expansions. In view of \eqref{eq: Z for N-contour} and \eqref{eq:
expansion Z la}, we thus have
\begin{equation}\label{eq: cl.exp. 01}
   \log Z_\ga^{\la,x_0,+} - \log Z_\ga^{\la,x_0,-} =
      2S^\la_{\La(\ga)} + 2\eta \mathbf{1}_{x_0\in \La(\ga)}
      + \sum_{C: \, C\sqsubset\Inn\ga}
    \De\Phi_C^{\la,x_0},
\end{equation}
where $\De\Phi_C^{\la,x_0} = \Phi_C^{\la,x_0,+} -
\Phi_C^{\la,x_0,-}$ and the sum runs only over the clusters $C$
such that all of its contours are in $\Inn\ga$. Observing that
$\De\Phi_C^{\la,x_0}$ vanishes whenever $\La(C) \cap \dL_n =
\emptyset$ or $\La(C)\not\ni x_0$, we get
\begin{equation}\label{eq: bound on surface free en.}
  \sum_{C: \, C\sqsubset\Inn\ga}
         \hspace{-2mm}  |\De\Phi_C^{\la,x_0}|
  \leq \sum_{y\in \dif_n(\ga)} \;
  \sum_{C: \, C\sqsubset\Inn\ga \atop \La(C)\ni y}
       |\De\Phi_C^{\la,x_0}| \,
  + \sum_{C: \, C\sqsubset\Inn\ga \atop \La(C)\ni x_0}
       |\De\Phi_C^{\la,x_0} |,
\end{equation}
where $\dif_n(\ga)=\La(\ga) \cap \dL_n$. Using the inductive
assumption \eqref{eq: phi-decay} and Lemma~\ref{Lem: geometrical},
the former sum may be estimated by
\begin{multline}
  \sum_{y\in \dif_n(\ga)} \; \sum_{C: \, \La(C)\ni y}
     |\De\Phi_C^{\la,x_0}| \leq 2 e^{-2(\be-c_1)}
     |\dif_n(\ga)| \leq \\
  \leq 2 e^{-2(\be-c_1)} |\cP(\ga)| \leq 2d e^{-2(\be-c_1)} |\ga|
       \leq |\ga|
\end{multline}
once $\tau\geq \tau_1 = c_1 + \frac12\, \log 2d$, while the latter
sum is smaller than $1$ if $\tau\geq c_1+\frac12\,\log 2$.
Combining these bounds with the definition \eqref{eq: K for
N-contour} of $K_\ga^{\la,x_0,\pm}$, we therefore find
\begin{equation}\label{eq: aux-K}
  K_\ga^{\la,x_0,\pm} \leq e^{-2 \left(\be|\ga| -
     |S^\la_{\La(\ga)}| - \eta^* - |\ga| - 1 \right)} \leq
  e^{-2(\be - \th\la^* - \eta^* - 2)|\ga|}
\end{equation}
as long as $\tau\geq\tau_1$. Moreover,
\begin{equation}\label{eq: aux-der.of K}
  \Bigl| \,\frac{\dif K_\ga^{\la,x_0,\pm}}{\dif\eta} \,\Bigr|
      \leq 4 K_\ga^{\la,x_0,\pm}
\end{equation}
for any $\tau\geq c_2$. To see this, it suffices to combine
\eqref{eq: K for N-contour} with the bound
\begin{equation}
  \Bigl| \, \frac{\dif}{\dif\eta}\, (\log Z_\ga^{\la,x_0,\mp} -
        \log Z_\ga^{\la,x_0,\pm}) \Bigr| \leq 2 +
  \sum_{C: \, C\sqsubset\Inn\ga \atop \La(C)\ni 0}  \Bigl| \,
  \frac{\dif\De\Phi_C^{\la,x_0}}{\dif\eta} \,\Bigr| \leq 4
\end{equation}
following from the cluster expansion \eqref{eq: cl.exp. 01} and
the inductive assumption \eqref{eq: phi'-decay}. Using \eqref{eq:
aux-K} and \eqref{eq: aux-der.of K}, the arguments from the case
$N=0$ readily yield the bounds \eqref{eq: phi-decay} to \eqref{eq:
K'-decay} if $c'_1 \geq \th\la^* + \eta^* + 2 + \frac12\, \log
2\vk$, $c_1 \geq c'_1+1+ \frac12\, \log (2.3^d\vk)$, $c'_2 \geq
\th\la^* + \eta^* + 2 + \frac12\, \log 2\vk$, $c_2 \geq c'_2 + 1 +
\frac12\, \log(4.3^d\vk)$, and $\tau\geq \max\{c_1 + \log 2d,\,
c_2\}$.
\end{proof}
\medz

\begin{proof}[Proof of Lemma~\ref{lem: cl.exp. +-}.]
The proof goes along the same lines as that of Lemma~\ref{Lem:
cl.exp.}. It should be clear that the constant $\hat \tau$ can be
chosen in such a way that $\hat\tau \leq \tau - \th\la^*$.
\end{proof}

\section{Proof of Lemma~\ref{lem: char.fct.}}
\label{sec: proof of Lemma 5.1}

In order to prove the uniformness in the distribution of the
boundary fields $\la$, we will need the following lemma.


\begin{lem}\label{lem: cl.exp.-rozdiely}
There exist constants $c_3 < \tilde\tau < \infty$ depending on $d$
and $\la^*$ such that for any $\be\geq\tilde\tau$ one has
\begin{equation}\label{eq: Dphi-decay}
   \sum_{C:\, \La(C)\ni x} e^{2(\be-c_3)|C|}
     \frac{ |\De\Phi^{\la}_C|} {
        \sinh\bigl( 4\sum_{y\in \dif_n(C)} |\la_y| \bigr) } \leq 1
\end{equation}
for any $x\in\La_n$ and $n\in\N$, provided $|\la_x|\leq\la^*$ for
all $x\in\Z^d$. Here $\De\Phi^\la_C=\Phi^{\la,+}_C -
\Phi^{\la,-}_C$, $\dif_n(C)=\cup_{\ga\in C} \dif_n(\ga)$ with
$\dif_n(\ga)=\La(\ga)\cap\dL_n$, and in the summation we adopt the
convention that $0.\infty=0$.
\end{lem}

\begin{proof}
Let $\be\geq\tilde\tau$ with $\tilde\tau$ to be determined and
$x\in\La_n$. Proceeding by induction on $N\in\N$, where $N$ is the
maximal size of $v(\ga)$ and $v(C)$ for $\ga$ and $C$ under
consideration, along with the estimate \eqref{eq: Dphi-decay} we
shall also prove that
\begin{equation}\label{eq: DK-decay}
   \sum_{\ga:\, \La(\ga)\ni x} e^{2(\be-c'_3)|\ga|}
     \frac{ |\De K^\la_\ga|} {
        \sinh\bigl( 4\sum_{y\in \dif_n(\ga)} |\la_y| \bigr) } \leq 1
\end{equation}
for some $c'_3<\tilde\tau$ (depending on $d$ and $\la^*$), where
$\De K^\la_\ga = K^{\la,+}_\ga - K^{\la,-}_\ga$.

First, let $N=0$. Then \eqref{eq: Z,K for 1-contour} yields
\begin{equation}
  |\De K^\la_\ga| \leq 2e^{-2\be|\ga|} \sinh \Bigl(
       2\sum_{x\in \dif_n (\ga)} |\la_x| \Bigr).
\end{equation}
As a consequence,
\begin{equation}
   \sum_{\ga:\, \La(\ga)\ni x \atop v(\ga)=0}
   \frac{ e^{2(\be-c'_3)|\ga|} \,
   |\De K_\ga^\la|}{\sinh \bigl(4\sum_{y\in \dif_n(\ga)}
      |\la_y| \bigr)} \leq
   2 \sum_{\ga:\, \La(\ga)\ni x \atop v(\ga)=0}
   e^{-2c'_3|\ga|} \leq  4 (\vk e^{-2c'_3})^d \leq 1
\end{equation}
once $c'_3 \geq \frac12\, \log 2\vk$, which verifies \eqref{eq:
DK-decay} and leads to the estimate
\begin{equation}
  \sum_{\ga'\io\ga:\, v(\ga')=0} \; \frac{e^{2(\be-c'_3)|\ga'|} \,
   |\De K_{\ga'}^\la|}{\sinh \bigl(4\sum_{x\in \dif_n(\ga')}
   |\la_x| \bigr)} \leq 3^d|\ga| .
\end{equation}
Using Corollary~\ref{cor: rozdiely}, the latter implies (\cf
\eqref{eq: aux-ref})
\begin{equation}
  \sum_{C:\, \La(C)\ni x \atop v(C)=0}
     \frac{e^{2(\be-c_3)|C|} \, |\De\Phi^{\la}_C|} {
     \sinh\bigl( 4\sum_{y\in \dif_n(C)} |\la_y| \bigr) }
  \leq 2 \sum_{\ga:\, \La(\ga)\ni x \atop v(\ga)=0}
     e^{-[2(c_3-c'_3-2)-d\log 3]\,|\ga|} \leq 1
\end{equation}
if $c'_3 \geq c'_1$, $\tilde\tau \geq \tilde\tau_1 = c'_3+1+ \frac
d2\,\log 3$, and $c_3\geq c'_3+2+ \frac12\, \log (2.3^d\vk)$ (here
$c'_1$ is the constant from \eqref{eq: K-decay}), which proves
\eqref{eq: Dphi-decay} in the case $N=0$.

Supposing now that the estimates \eqref{eq: Dphi-decay} and
\eqref{eq: DK-decay} have been proved for all integers smaller
than $N$, let us prove them for $N$. The relations \eqref{eq: K
for N-contour} and \eqref{eq: cl.exp. 01} with $\eta=0$ yield
\begin{equation}\label{eq: aux-DK bound}
  |\De K^\la_\ga| \leq 2e^{-2\be|\ga|} \sinh \Bigl(
  2\sum_{x\in \dif_n(\ga)} |\la_x| +
  \sum_{C:\, C \sqsubset \Inn\ga \atop \dif_n(C) \neq\emptyset}
  |\De\Phi_C^\la| \Bigr),
\end{equation}
where in the second sum only the clusters $C$ such that all of
their contours lie in $\Inn\ga$ are considered. Using the
inductive assumption \eqref{eq: Dphi-decay}, let us first show
that
\begin{equation}\label{eq: aux-clasters vs S}
  \sum_{C:\, C \sqsubset \Inn\ga \atop \dif_n(C)\neq\emptyset}
  |\De\Phi_C^\la|  \leq 2\sum_{x\in \dif_n(\ga)} |\la_x|
\end{equation}
whenever $\tilde\tau \geq \tilde\tau_2 = c_3 + 2\th\la^* + 1 +
\frac12\, \log 4\th$. Since $\sinh x \leq x e^x$ for any $x\geq
0$, the left-hand side of the last inequality can be bounded by
\begin{multline}
  4e^{-2(\be-c_3-2\th\la^*-1)} \sum_{x\in \dif_n(\ga)} \;
  \sum_{C:\, C\sqsubset\Inn\ga \atop \La(C)\ni x}
      \frac{e^{2(\be-c_3-2\th\la^*-1)|C|}\,
  |\De\Phi_C^\la|}{\sinh \bigl(4\sum_{y\in \dif_n(C)}
      |\la_y| \bigr)} \; \times \\
  \times \; e^{4\th\la^*|C|} \sum_{z\in \dif_n(C)} |\la_z| \leq \\
  \leq \frac1\th \sum_{z\in \dif_n(\ga)}  |\la_z|
      \sum_{C:\, \La(C)\ni z} \; \sum_{x\in \dif_n(C)}
       \frac{e^{2(\be-c_3-1)|C|}\, |\De\Phi_C^\la|}{\sinh
      \bigl(4\sum_{y\in \dif_n(C)} |\la_y| \bigr)} \,.
\end{multline}
Realizing that the last summand is independent of $x$ and that one
has $|\dif_n(C)| \leq \sum_{\ga'\in C} |\dif_n(\ga')| \leq
\th|C|$, we obtain \eqref{eq: aux-clasters vs S}. Combining this
estimate and \eqref{eq: aux-DK bound} with the arguments from the
case $N=0$, one readily arrives at \eqref{eq: Dphi-decay} and
\eqref{eq: DK-decay} for any $c'_3 \geq \max\{\frac12\, \log
2\vk,\, c'_1\}$, $c_3\geq c'_3+ 2 + \frac12\, \log (2.3^d\vk)$,
and $\tilde\tau \geq \max\{\tilde\tau_1,\, \tilde\tau_2\}=
\tilde\tau_2$.
\end{proof}

We are now ready to prove Lemma~\ref{lem: char.fct.}.


\begin{proof}[Proof of Lemma~\ref{lem: char.fct.}.]
Let $\be \geq \be_0$ and $|t| \leq \ep$, both the constants being
specified in the course of the proof. For any set $P$ of
plaquettes in $V_n$, let
\begin{equation}
  |P|_{\co} = \inf_{P' \supset P \atop \text{connected}} |P'|\ ,
\end{equation}
where the infimum is taken over all connected sets of plaquettes
containing $P$. It suffices to prove that
\begin{equation}\label{eq: ch.f.-to prove}
  \sum_{P \ni p} e^{\frac{\si^2 t^2}{4d} |P|_{\co}}
  |w_{P}(t)| \leq \frac{\si^2 t^2}{4d}
\end{equation}
holds for any plaquette $p \in \dif V_n$. Indeed, the last
inequality implies that the condition \eqref{eq: KP-general} is
satisfied with $a(P) = \frac{\si^2 t^2}{4d} |P|_{\co}$ and $b(P) =
0$. Considering the polymer $\dif V_n$ and realizing that $|\dif
V_n|_{\co} \leq 2d|\dL_n|$, Proposition~\ref{prop: cl.exp.}
immediately yields \eqref{eq: bound on cl.}.

So, let us prove \eqref{eq: ch.f.-to prove}. Recalling that
$\De\Phi_C^\la$ is an odd function of $\la$, we may use the
symmetry of the distribution $\Pro$ to cast \eqref{eq: vr} into a
more suitable form, namely,
\begin{multline}
  \vr_{\mfr C}(t) = (\vp(2t))^{-|\dif_n(\mfr C)|}\;
  \Exp\, \Bigl\{\begin{array}{l}
                  \sin \\ \cos
                \end{array} \Bigr\}
  \Bigl[ t \Bigl(2S_{\dif_n(\mfr{C})}^\la + \frac{1}{2}
  \sum_{C\in\mfr C} \De\Phi_C^\la \Bigr) \Bigr] \times
\\
  \times \prod_{C\in\mfr C} 2\sin \Bigl(
  \frac{t \De\Phi_C^\la}{2} \Bigr)\ .
\end{multline}
Here $\sin$ is to be taken iff the cardinality of the set of
clusters $\card(\mfr C)$ is odd and $\cos$ whenever the
cardinality is even; to distinguish both cases, we will use the
notation $r_{\mfr C} = 1$ and $r_{\mfr C} = 0$, respectively.
Taking now $\ep$ such that $\ep\la^* \leq \frac12$, one has
$\vp(2t) = \Exp\, \cos(2t\la_0) \geq \frac{1}{2}$ whenever $|t|
\leq \ep$, and we can estimate
\begin{equation}
  |\vr_{\mfr C}(t)| \leq 2^{|\dif_n(\mfr C)|}
  t^{\card(\frC) + r_{\frC}} \;\Exp\,
  \biggl( 2|S_{\dif_n(\frC)}| + \frac{1}{2} \sum_{C \in \frC}
  |\De\Phi_C^\la| \biggr)^{r_{\frC}} \prod_{C \in \frC}
  |\De\Phi_C^\la|\ .
\end{equation}
Since $|\la_x| \leq \la^*$ and $\sinh x \leq e^x\min\{1,x\}$ for
any $x \geq 0$, with the help of Lemma~\ref{Lem: geometrical} we
have the inequality
\begin{equation}\label{eq: jedna}
  |\De\Phi_C^\la| \leq z_{C}^{\la} \min\{1,\, 4\sum_{y \in \dif_n(C)}
     |\la_y|\}\ ,
\end{equation}
where we introduced the shorthand
\begin{equation}
  z_{C}^{\la} = \frac{|\De\Phi_{C}^{\la}|}{\sinh (4\sum_{x\in\dif_n(C)}
  |\la_x|)} \ e^{4\th\la^*|C|}\ .
\end{equation}
Lemma~\ref{lem: cl.exp.-rozdiely} then gives the estimate
\begin{equation}\label{eq: dva}
\begin{split}
  \frac{1}{2} \sum_{C \in \frC} |\De\Phi_C^\la| &\leq
  2\sum_{y \in \dif_n(\frC)} |\la_y| \sum_{C\in\frC \atop y \in \dif_n(C)}
  z_{C}^{\la} \leq
\\
  &\leq 2\sum_{y \in \dif_n(\frC)} |\la_y| \sum_{C:\, y \in \La(C)}
  z_{C}^{\la} \leq 2\sum_{x \in \dif_n(\frC)} |\la_x|
\end{split}
\end{equation}
provided $\be_0 \geq 2\th\la^* + c_3$, where $c_3$ is the constant
from Lemma~\ref{lem: cl.exp.-rozdiely}. Using the bounds
\eqref{eq: jedna} and \eqref{eq: dva}, we obtain
\begin{equation}
\begin{split}
  |\vr_{\mfr C}(t)| &\leq 2^{|\dif_n(\mfr C)|+1}\,
  |t|^{\card(\frC) + r_{\frC}} \;\Exp\, \biggl( 4\sum_{x\in\dif_n(\frC)}
  |\la_x| \biggr) \times
\\
  &\hspace{20mm} \times \prod_{C\in\frC} z_{C}^{\la}
  \min\{1,4\sum_{x\in\dif_n(C)} |\la_x|\} \leq
\\
  &\leq  2^{|\dif_n(\mfr C)|+1}\, t^2 \;
  \Exp\, \biggl( 4\sum_{x\in\dif_n(\frC)} |\la_x| \biggr)^2
  \prod_{C\in\frC} z_{C}^{\la}
\end{split}
\end{equation}
once $\ep \leq 1$. Since $\Exp\, |\la_x| \leq (\Exp\,
\la_x^2)^{1/2}$ by the Cauchy-Schwartz inequality, we have
\begin{equation}
  \Exp\, \biggl( \sum_{x\in\dif_n(\frC)} |\la_x| \biggr)^2 =
  \sum_{x\in\dif_n(\frC)} \Exp\, \la_x^2 +
  \sum_{x,y\in\dif_n(\frC) \atop x \neq y} \Exp\, |\la_x|\;
  \Exp |\la_y| \leq \si^2 |\dif_n(\frC)|^2\ ,
\end{equation}
yielding the bound
\begin{equation}
  |\vr_{\frC}(t)| \leq 32 \si^2 t^2 \, 2^{|\dif_n(\frC)|} |\dif_n(\frC)|^2
  \prod_{C\in\frC} z_{C}^{\la} \leq 32\si^2 t^2 \prod_{C\in\frC}
  (2e^2)^{\th |C|} z_{C}^{\la} \ .
\end{equation}
Observing that, in view of Lemma~\ref{Lem: geometrical},
\begin{equation}
  |P|_{\co} \leq | P \cup (\cup_{C\in\frC_{\co}} C) | \leq
  (d+1) \sum_{C\in\frC_{\co}} |C|
\end{equation}
for any $\frC_{\co}$ with $\cP(\frC_{\co})=P$, we thus get
\begin{align}
 |w_P (t)| &\leq  32 \si^2 t^2 e^{-2\frac{\be-\vt}{d+1} |P|_{\co}}
   \sum_{{\mfr C}_{\co}: \atop \cP({\mfr C}_{\co}) = P} \;
   \prod_{C\in\frC} \tilde z_{C}^{\la} \leq \non \\
 &\leq 32 \si^2 t^2 e^{-2\frac{\be-\vt}{d+1}|P|_{\co}}
   \sum_{{\mfr C}:\, \cP(C) \cap P \neq \emptyset \atop
   \text{for all $C\in\frC$}} \;
   \prod_{C\in\frC} \tilde z_{C}^{\la}
\end{align}
for any $\vt>0$, where $\tilde z_C^\la = (2e^2)^{\th |C|}
e^{2(\be-\vt)|C|} z_{C}^{\la}$. With the help of Lemma~\ref{lem:
cl.exp.-rozdiely}, the last sum may be further estimated by
\begin{equation}
 \prod_{C:\, \cP(C) \cap P \neq\emptyset} (1 + \tilde z_{C}^{\la})
  \leq e^{\sum_{C:\, \cP(C) \cap P \neq\emptyset}
    \tilde z_{C}^{\la}}
  \leq e^{|P|\,\inf_x \sum_{C:\, x\in\La(C)} \tilde z_{C}^{\la}}
  \leq e^{|P|}
\end{equation}
whenever $\vt \geq 2\th\la^* + c_3 + d(1 + \frac{1}{2}\ln 2)$.
Realizing that $\si^2 t^2 \leq (\la^* \ep)^2 \leq 1$ and that the
number of connected polymers containing a given plaquette and
having the size $|P|=l$ may be estimated by $\ka^{l}$ with a
constant $\ka
> 0$, we finally get
\begin{equation}
\begin{split}
  \sum_{P \ni p} e^{\frac{\si^2 t^2}{4d} |P|_{\co}}
    &|w_{P}(t)| \leq
  32\si^2 t^2 \sum_{P \ni p} \sum_{P' \supset P \atop \text{connected}}
    e^{-2\left( \frac{\be-\vt}{d+1} - 1 \right)|P'|} \leq
\\
  &\leq 32\si^2 t^2 \sum_{P' \ni p \atop \text{connected}}
    \Bigl[ 2 e^{ -2\left( \frac{\be-\vt}{d+1} - 1 \right)} \Bigr]^{|P'|} \leq
\\
  &\leq 32\si^2 t^2 \sum_{l=1}^\infty
     \Bigl[ 2\ka e^{ -2\left( \frac{\be-\vt}{d+1} - 1 \right)} \Bigr]^l \leq
  \frac{\si^2 t^2}{4d} \ ,
\end{split}
\end{equation}
providing that $\be_0 \geq \vt + (d+1)[1+4\ln 2 + \frac12\,
\log(2d\ka)]$, say.
\end{proof}

\section*{Acknowledgements}

Both I.~M.~and K.~N.~would like to thank A.~C.~D.~van Enter and
express their appreciation for the hospitality extended to them
during their stays at the Institute for Theoretical Physics of
Groningen University. I.~M.~also thanks C.~Maes and gratefully
acknowledges the hospitality he received during his visit to the
Institute for Theoretical Physics at the Catholic University of
Leuven. K.~N.~ is grateful to C.~Maes for very instructive discussions.
Also a discussion with R. Koteck\'y is gratefully acknowledged.


\bibliographystyle{plain}

\begin{thebibliography}{10}

\bibitem{BI89}
C.~Borgs and J.~Z. Imbrie.
\newblock A {Unified} {Approach} to {Phase} {Diagrams} in {Field} {Theory} and
  {Statistical} {Mechanics}.
\newblock {\em Commun. Math. Phys.}, 123:305--328, 1989.

\bibitem{BK90}
C.~Borgs and R.~Koteck\'y.
\newblock A {Rigorous} {Theory} of {Finite}-{Size} {Scaling} at {First}-{Order}
  {Phase} {Transitions}.
\newblock {\em Journal of Stat. Phys.}, 61:79--119, 1990.

\bibitem{BK95}
C.~Borgs and R.~Koteck\'y.
\newblock Surface-{Induced} {Finite}-{Size} {Effects} for the {First}-{Order}
  {Phase} {Transition}.
\newblock {\em J. Stat. Phys.}, 79:43--116, 1995.

\bibitem{BKM91}
C.~Borgs, R.~Koteck\'y, and S.~{Miracle-Sol\'e}.
\newblock Finite-{Size} {Scaling} for the {Potts} {Models}.
\newblock {\em Journal of Stat. Phys.}, 62:529--552, 1991.

\bibitem{BovGay}
A.~Bovier and V.~Gayrard.
\newblock {Hopfield} {Models} as {Generalized} {Random} {Mean}-{Field}
  {Models}.
\newblock In A.~Bovier and P.~Picco, editors, {\em {Mathematical} {Aspects} of
  {Spin} {Glasses} and {Neural} {Networks}}, pages 3--89. Birkh{\"a}user,
Boston 1998.

\bibitem{Boven}
A.~Bovier, A.~C.~D. {van Enter}, and B.~Niederhauser.
\newblock {Stochastic} {Symmetry} {Breaking} in a {Gaussian} {Hopfield}
  {Model}.
\newblock {\em J. Stat. Phys.}, 95:181--213, 1999.
\newblock See also B. Niederhauser's thesis ``Mathematical Aspects of Hopfield
  Models'' (2000) at www.math.tu-berlin.de/stoch/Kolleg/homepages/Niederhauser.

\bibitem{BoZa}
A.~Bovier and M.~Zahradn{\ii}k.
\newblock A {Simple} {Inductive} {Approach} to the {Problem} of {Convergence}
  of {Cluster} {Expansions} of {Polymer} {Models}.
\newblock {\em J. Stat. Phys.}, 100:765--778, 2000.

\bibitem{Campent}
M.~Campanino and A.~C.~D. {van Enter}.
\newblock {Weak} {Versus} {Strong} {Uniqueness} of {Gibbs} {Measures}: {A}
  {Regular} {Short}-{Range} {Example}.
\newblock {\em J. Phys. A, Math. Gen.}, 28:L45--L47, 1995.

\bibitem{Du}
R.~Durrett.
\newblock {\em Probability: {Theory} and {Examples}}.
\newblock Wadsworth, Inc., Belmont, 1985.

\bibitem{FH}
D.~S. Fisher and D.~A. Huse.
\newblock {Pure} {States} in {Spin} {Glasses}.
\newblock {\em J. Phys. A, Math. and Gen.}, 20:L997--L1003, 1987.

\bibitem{Ge}
H.~O. Georgii.
\newblock {\em Gibbs {Measures} and {Phase} {Transitions}}.
\newblock de Gruyter, Berlin, 1988.

\bibitem{Higyos}
Y.~Higuchi and N.~Yoshida.
\newblock Slow {Relaxation} of {2D} {Stochastic} {Ising} {Models} with {Random}
  and {Non}-{Random} {Boundary} {Conditions}.
\newblock In K.~Elworthy, S.~Kusuoka, and I.~Shikegawa, editors, {\em New
  Trends in Stochastic Analysis}, pages 153--167. World Scientific, Singapore,
  1997.

\bibitem{KoPr}
R.~Koteck\'y and D.~Preiss.
\newblock Cluster {Expansions} for {Abstract} {Polymer} {Models}.
\newblock {\em Commun. Math. Phys.}, 103:491--498, 1986.

\bibitem{Kue97}
C.~K{\"u}lske.
\newblock Metastates in {Disordered} {Mean}-{Field} {Models}: {Random} {Field}
  and {Hopfield} {Models}.
\newblock {\em J. Stat. Phys.}, 88:1257--1293, 1997.

\bibitem{Kue98pr}
C.~K{\"u}lske.
\newblock Limiting {Behavior} of {Random} {Gibbs} {Measures}: {Metastates} in
  {Some} {Disordered} {Mean}-{Field} {Models}.
\newblock In A.~Bovier and P.~Picco, editors, {\em Mathematical Aspects of Spin
  Glasses and Neural Networks}, pages 151--160. Birkh{\"a}user, Boston, 1998.

\bibitem{Kue98}
C.~K{\"u}lske.
\newblock Metastates in {Disordered} {Mean}-{Field} {Models} ii.
\newblock {\em J. Stat. Phys.}, 91:155--176, 1998.

\bibitem{LebPen}
J.~L.~Lebowitz and O.~Penrose.
\newblock Thermodynamic Limit of the Free Energy and Correlation
Functions of Spin Systems.
\newblock Acta Physica Austriaca, Suppl. XVI: 201--220, 1976.

\bibitem{Li}
T.~M. Liggett.
\newblock {\em Interacting {Particle} {Systems}}.
\newblock Springer-Verlag, Berlin, 1985.


\bibitem{Med01}
I. Medve\md.
\newblock Finite-{Size} {Effects} for {Classical} {Lattice} {Models}.
\newblock Ph.D. thesis, Charles University, Prague, 2001.
\newblock See also C.~Borgs, R.~Koteck\'y, and I.~Medve\md. Finite-{Size}
   {Effects} for the {Potts} {Model} with {Weak} {Boundary} {Conditions}.
   Submitted to J.~Stat.~Phys.
\newblock A preliminary version in Igor Medve\md. Finite-{Size} {Effects}
   for the {Potts} {Model}. Master's thesis, Charles University, Prague, 1996.

\bibitem{MS00}
S.~{Miracle-Sol\'e}.
\newblock On the {Convergence} of {Cluster} {Expansions}.
\newblock {\em Physica A}, 279:244--249, 2000.

\bibitem{Ne97}
C.~M. Newman.
\newblock {\em Topics in {Disordered} {Systems}}.
\newblock Lectures in Mathematics ETH-Z{\"u}rich. Birkh{\"a}user, Basel, 1997.

\bibitem{NeSt92}
C.~M. Newman and D.~L. Stein.
\newblock Multiple {States} and {Thermodynamic} {Limits} in {Short}-{Ranged}
  {Ising} {Spin}-{Glass}.
\newblock {\em Phys. Rev. B}, 46:973--982, 1992.

\bibitem{NeSt97}
C.~M. Newman and D.~L. Stein.
\newblock Metastate {Approach} to {Thermodynamic} {Chaos}.
\newblock {\em Phys. Rev. E}, 55:5194--5211, 1997.

\bibitem{NeSt98}
C.~M. Newman and D.~L. Stein.
\newblock {Thermodynamic} {Chaos} and the {Structure} of {Short}-{Range} {Spin}
  {Glasses}.
\newblock In A.~Bovier and P.~Picco, editors, {\em {Mathematical} {Aspects} of
  {Spin} {Glasses} and {Neural} {Networks}}, pages 243--287. Birkh{\"a}user,
Boston, 1998.

\bibitem{NeSt02}
C.~M. Newman and D.~L. Stein.
\newblock The state(s) of replica symmetry breaking: Mean field theories
  vs.~short-ranged spin glasses.
\newblock {\em J. Stat. Phys.}, 106:213--244, 2002.
\newblock Formerly known as: ``Replica Symmetry Breaking's New Clothes''.

\bibitem{Sok01}
A.~D. Sokal.
\newblock Chromatic {Polynomials}, {Potts} {Models}, and {All} {That}.
\newblock {\em Physica A}, 279:324-332, 2000.

\bibitem{vE90}
A.~C.~D. van Enter.
\newblock {Stiffness} {Exponent}, {Number} of {Pure} {States} and
  {Almeida}-{Thouless} {Line} in {Spin}-{Glasses}.
\newblock {\em J. Stat. Phys.}, 60:275--279, 1990.

\bibitem{vE00}
A.~C.~D. van Enter.
A Remark on the Notion of Robust Phase Transitions
\newblock {\em J. Stat. Phys.}, 98:1409--1416, 2000.

\bibitem{vES02}
A.~C.~D. van Enter and H.~G. Schaap.
\newblock Infinitely many states and stochastic symmetry in a gaussian
  potts-hopfield model.
\newblock {\em J. Phys. A}, 2002.
\newblock To appear.

\bibitem{Z84}
M.~Zahradn{\ii}k.
\newblock An {Alternate} {Version} of {Pirogov}-{Sinai} {Theory}.
\newblock {\em Commun. Math. Phys.}, 93:559--581, 1984.

\end{thebibliography}
\end{document}